\documentclass[onecolumn,superscriptaddress]{revtex4-2}

\usepackage{amsmath}
\usepackage{amssymb}
\usepackage{bm}
\usepackage{graphicx}
\usepackage{caption}
\usepackage{subcaption}
\usepackage{booktabs}
\usepackage{mathtools}
\usepackage{physics}

\usepackage{braket,tensor,slashed,comment,mathtools}
\usepackage{tabularx, tikz-cd}

\usepackage{pdfpages}

\makeatletter
\AtBeginDocument{\let\LS@rot\@undefined}
\makeatother

\usepackage[
    colorlinks=true,
    linkcolor=blue,
    citecolor=blue,
    urlcolor=blue
]{hyperref}

\makeatletter
\AtBeginDocument{%
\long\def\@makecaption#1#2{%
  \par
  \vskip\abovecaptionskip
  \begingroup
    \small\rmfamily
    \samepage
    \flushing
    \let\footnote\@footnotemark@gobble
    \@make@capt@title{#1}{#2}\par
  \endgroup
  \vskip\belowcaptionskip
}%
}
\makeatother

\renewcommand{\vec}[1]{{\boldsymbol #1}}

\begin{document}

\foreach \p in {1,...,15}{%
    \clearpage
    \includepdf[
        pages={\p},
        fitpaper=true,
        pagecommand={\thispagestyle{empty}}
    ]{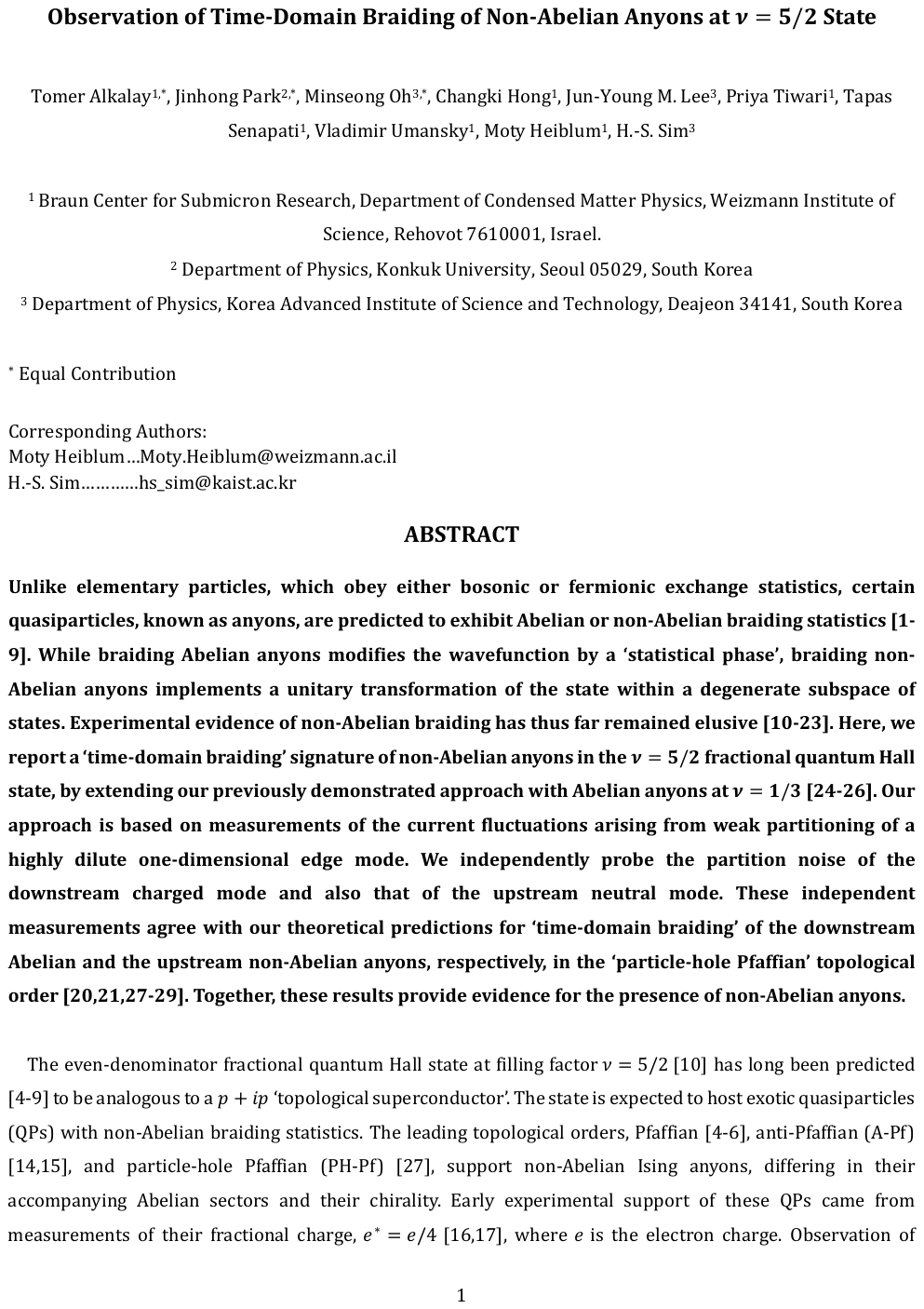}
}


\clearpage

\title{Supplementary Notes for ``Observation of Time-Domain Braiding of Non-Abelian Anyons at $\nu = 5/2$ State''}
\author{Tomer Alkalay} \email{equal contribution}
\affiliation{Braun Center for Submicron Research, Department of Condensed Matter Physics, Weizmann Institute of Science, Rehovot 7610001, Israel.}
\author{Jinhong Park} \email{equal contribution}
\affiliation{Department of Physics, Konkuk University, Seoul 05029, South Korea}
\author{Minseong Oh} \email{equal contribution}
\affiliation{Department of Physics, Korea Advanced Institute of Science and Technology, Daejeon 34141, South Korea}
\author{Changki Hong}
\affiliation{Braun Center for Submicron Research, Department of Condensed Matter Physics, Weizmann Institute of Science, Rehovot 7610001, Israel.}
\author{Jun-Young M. Lee}
\affiliation{Department of Physics, Korea Advanced Institute of Science and Technology, Daejeon 34141, South Korea}
\author{Priya Tiwari}
\affiliation{Braun Center for Submicron Research, Department of Condensed Matter Physics, Weizmann Institute of Science, Rehovot 7610001, Israel.}
\author{Tapas Senapati}
\affiliation{Braun Center for Submicron Research, Department of Condensed Matter Physics, Weizmann Institute of Science, Rehovot 7610001, Israel.}
\author{Vladmir Umansky}
\affiliation{Braun Center for Submicron Research, Department of Condensed Matter Physics, Weizmann Institute of Science, Rehovot 7610001, Israel.}
\author{Moty Heiblum} \email{Moty.Heiblum@weizmann.ac.il}
\affiliation{Braun Center for Submicron Research, Department of Condensed Matter Physics, Weizmann Institute of Science, Rehovot 7610001, Israel.}
\author{H.-S. Sim} \email{hs\_sim@kaist.ac.kr}
\affiliation{Department of Physics, Korea Advanced Institute of Science and Technology, Daejeon 34141, South Korea}
\date{\today}
	 
\maketitle

\section{Experimental data sets} \label{sec:comparison}
 
\begin{figure}[!htbp]
	\centering
	
	\includegraphics[
	width= \textwidth,
	height=0.85\textheight,
	keepaspectratio
	]{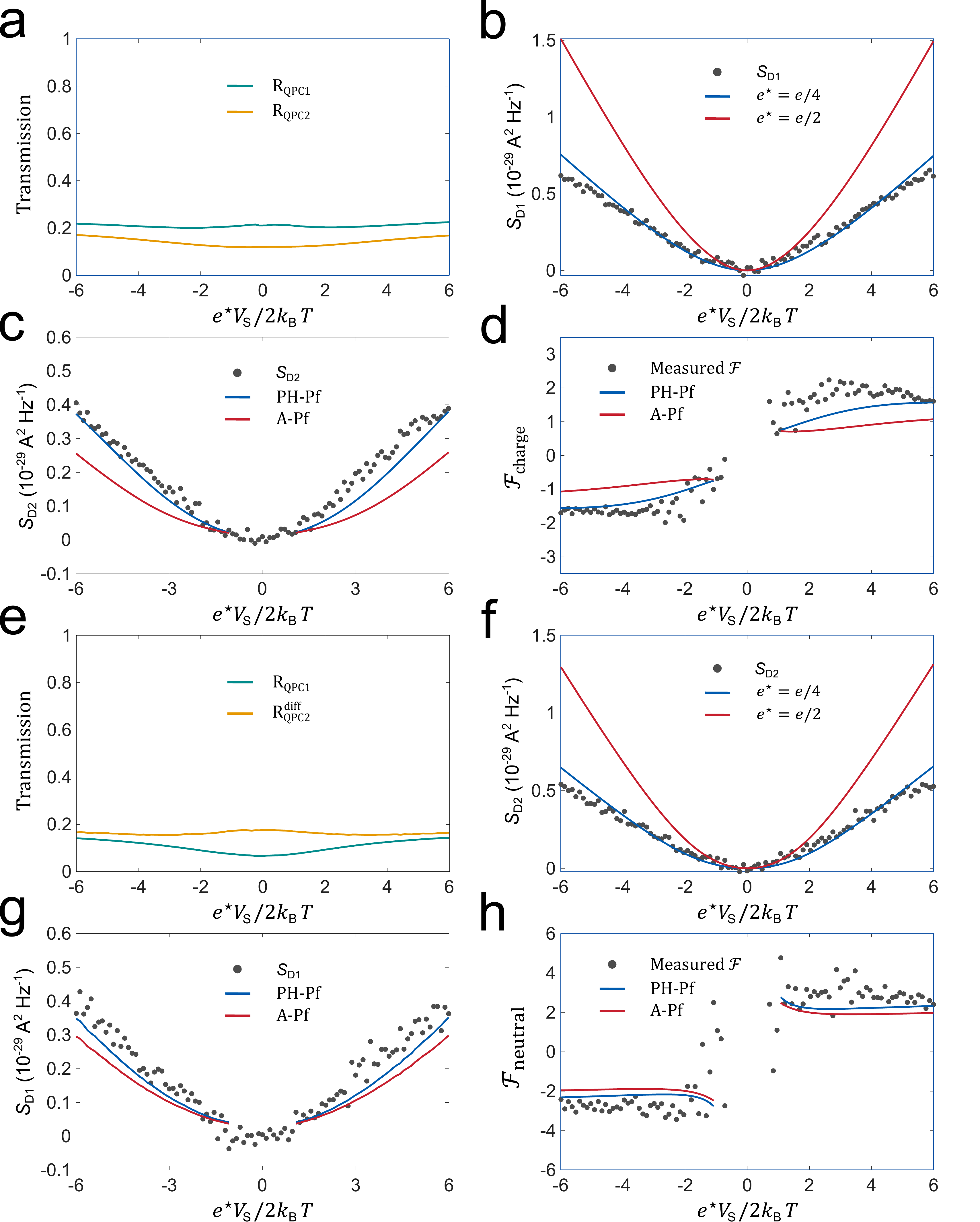}

	\caption{{\bf  Measurement data set 1 obtained from Device A.} Panels (a)–(d) show measurements in the charged-anyon configuration illustrated in Fig.~1(a) of the main text, whereas panels (e)–(h) show measurements in the neutral-anyon configuration shown in Fig.~3(a). All quantities are plotted as a function of $e^\ast V_{\rm S}/(2k_{\rm B}T)$.
    {\bf (a,e)} QPC reflection probabilities. Panel (a) shows $R_{\rm QPC1}$ and $R_{\rm QPC2}$ in the charged-anyon configuration, while panel (e) presents $R_{\rm QPC1}$ and $R_{\rm QPC2}^{\rm diff}$ in the neutral-anyon configuration. {\bf (b,f)} Fractional-charge measurement. The shot-noise spectral density generated by partitioning at the injection QPC of each configuration is shown by the gray dots and is measured at Drain D1 in panel (b) and Drain D2 in panel (f). The data are compared with the expected noise for quasiparticle charges $e^* = e/4$ (blue curve) and $e/2$ (red), based on Eq.~(1) of the main text. {\bf (c,g)} Time-domain braiding measurement. The excess noise generated at the detection QPC of each configuration is shown by the gray dots and is measured at Drain D2 in panel (c) and Drain D1 in panel (g). The data are compared with the theoretical predictions for the PH-Pf (blue curve) and A-Pf (red curve) edge structures. {\bf (d,h)} Experimental Fano factors extracted from the measured noise: $\mathcal{F}_{\rm charge}$ in panel (d) and $\mathcal{F}_{\rm neutral}$ in panel (h), as defined in Eqs.~(3) and (4) of the main text. The data are compared with the theoretical Fano factors predicted for the PH-Pf (blue curve) and A-Pf (red curve) edges.
    }

	\label{fig:S1}
\end{figure}

\begin{figure}[!htbp]
	\centering
	
	\includegraphics[
	width= \textwidth,
	height=0.85\textheight,
	keepaspectratio
	]{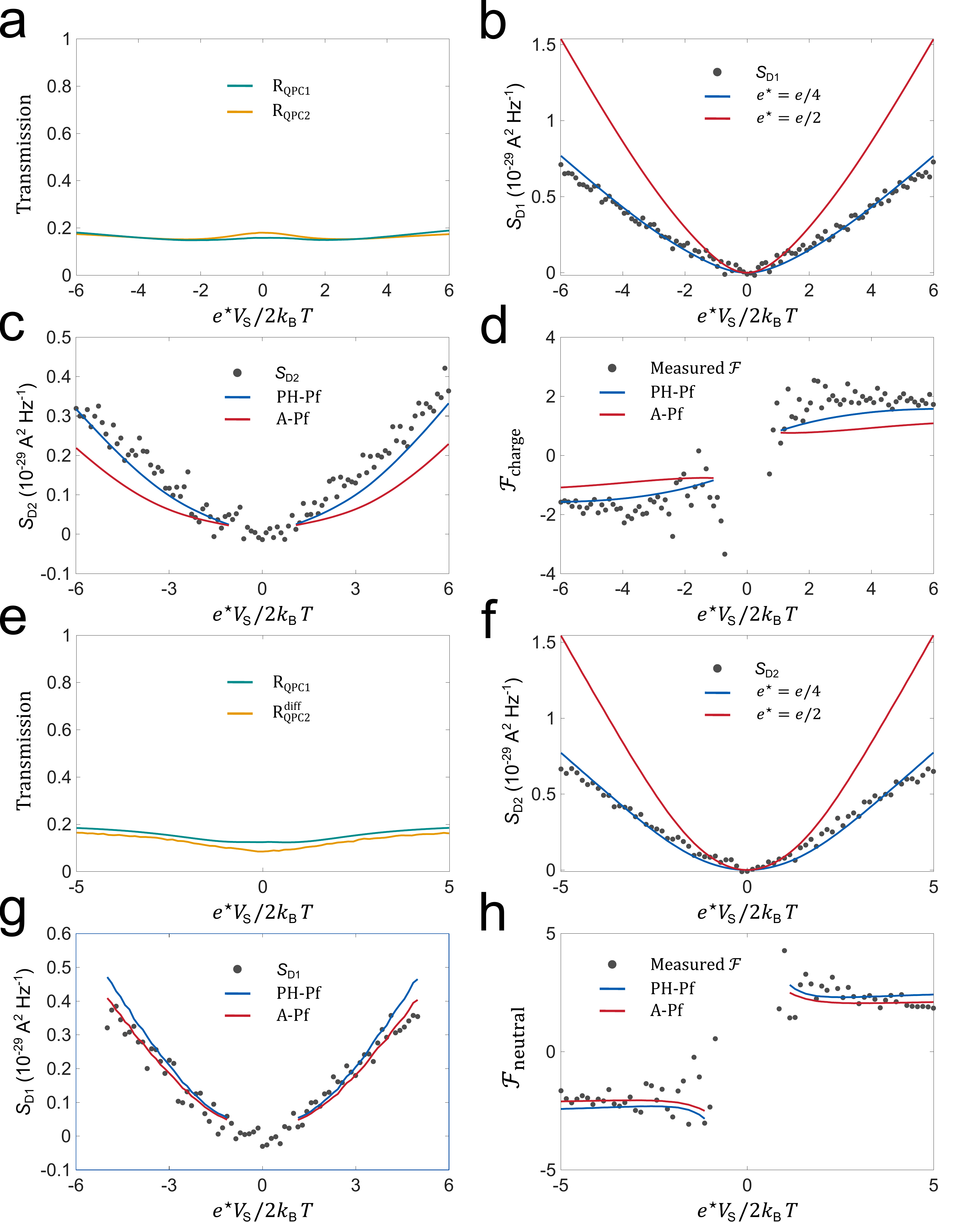}

	\caption{{\bf  Measurement data set 2 obtained from Device B.} The panels follow the same arrangement and notation as in Fig.~S1. Panels (a)–(d) show measurements in the charged-anyon configuration, while panels (e)–(h) show measurements in the neutral-anyon configuration. The data are compared with the corresponding fractional-charge and time-domain braiding predictions, as described in the caption of Fig.~S1.
    }

	\label{fig:S2}
\end{figure}

\begin{figure}[!htbp]
	\centering
	
	\includegraphics[
	width= \textwidth,
	height=0.85\textheight,
	keepaspectratio
	]{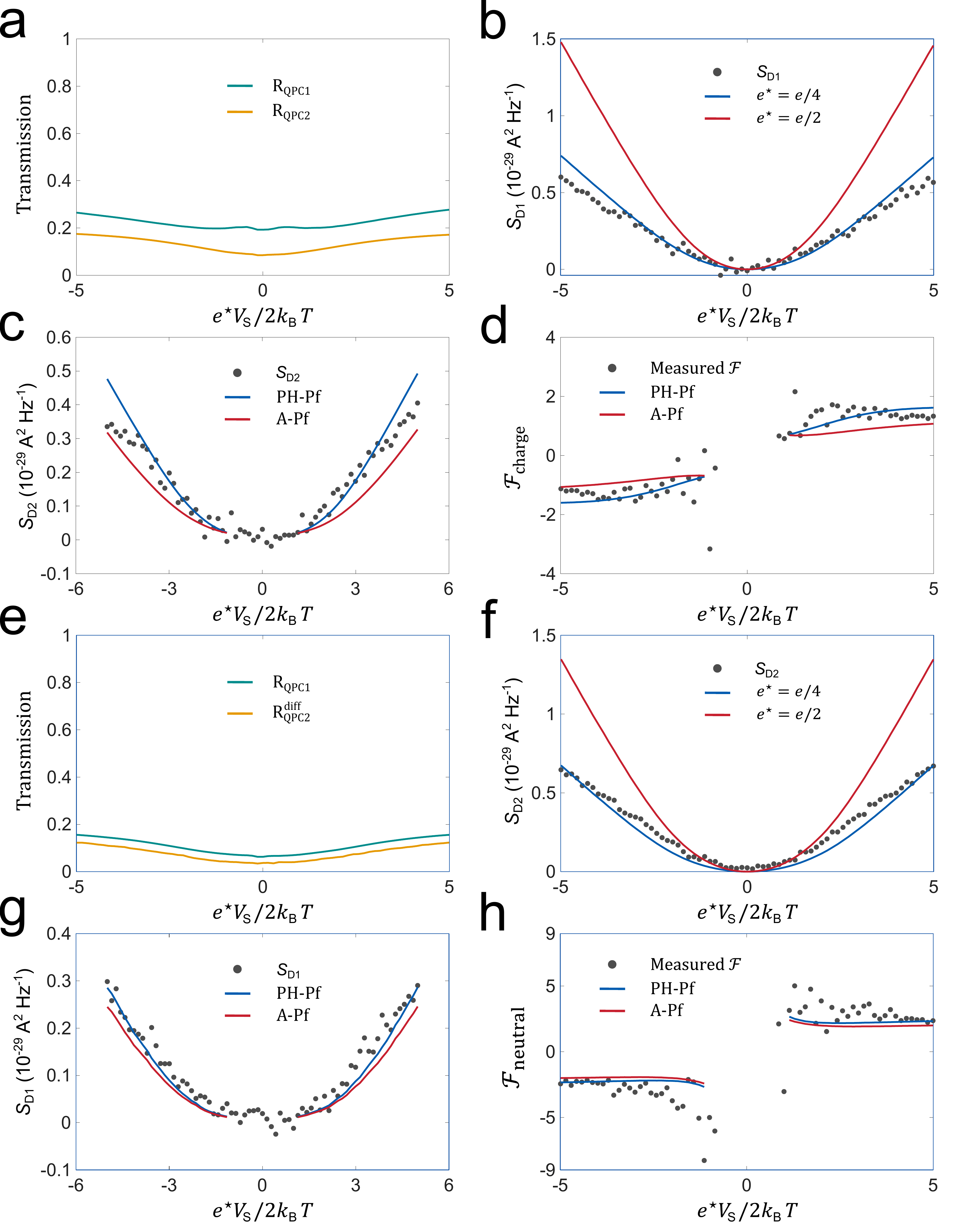}

	\caption{{\bf  Measurement data set 3 obtained from Device C.} The panels follow the same arrangement and notation as in Fig.~S1. Panels (a)–(d) show measurements in the charged-anyon configuration, while panels (e)–(h) show measurements in the neutral-anyon configuration. The data are compared with the corresponding fractional-charge and time-domain braiding predictions, as described in the caption of Fig.~S1.
    }

	\label{fig:S3}
\end{figure}

\begin{figure}[!htbp]
	\centering
	
	\includegraphics[
	width= \textwidth,
	height=0.85\textheight,
	keepaspectratio
	]{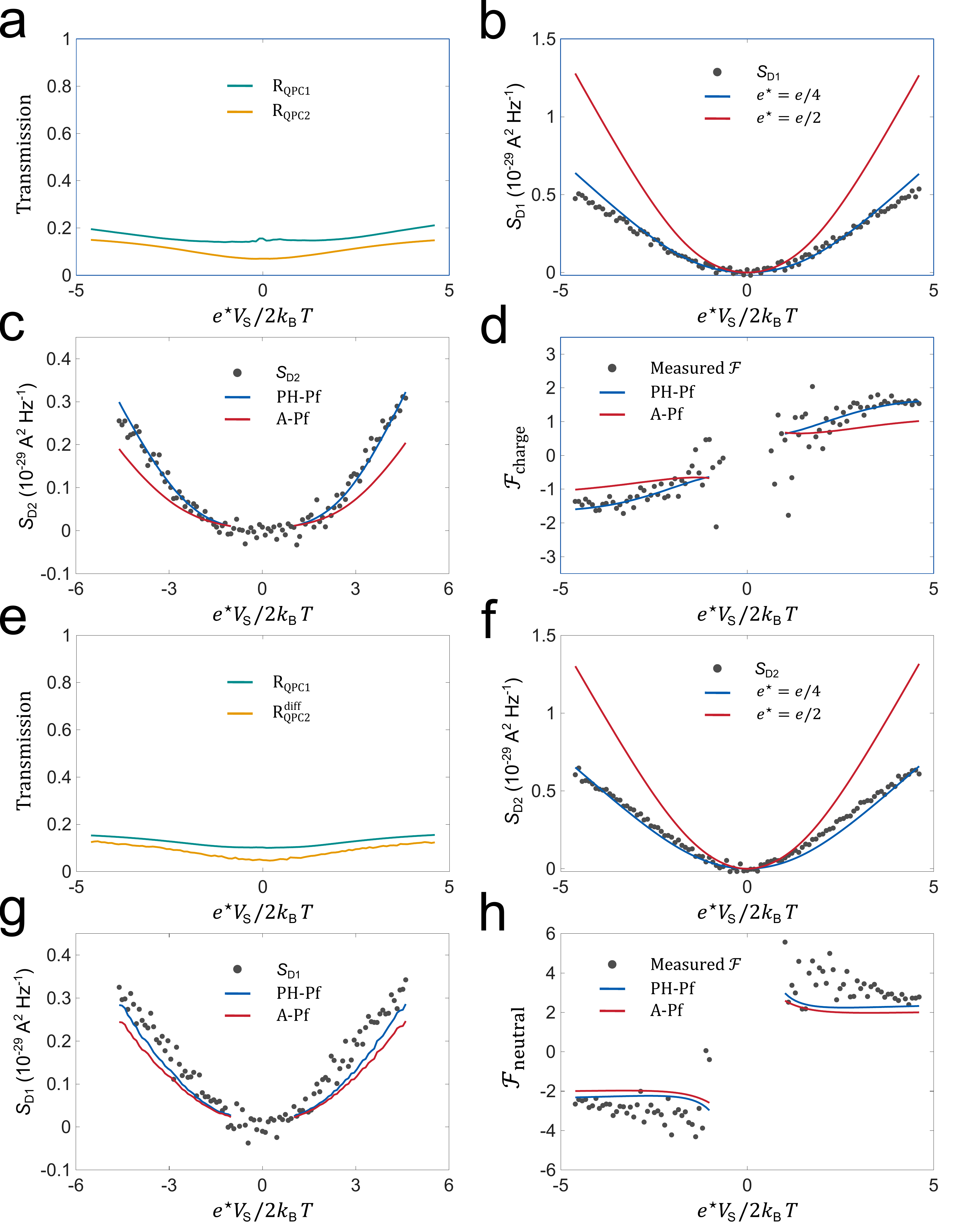}

	\caption{{\bf  Measurement data set 4 obtained from Device C.} The data were obtained during a separate cooldown from that of Fig.~S3, with a positive gate voltage of $0.3$~V applied during the cooldown. The panels follow the same arrangement and notation as in Fig.~S1. Panels (a)–(d) show measurements in the charged-anyon configuration, while panels (e)–(h) show measurements in the neutral-anyon configuration. The data are compared with the corresponding fractional-charge and time-domain braiding predictions, as described in the caption of Fig.~S1.
    }

	\label{fig:S4}
\end{figure}

Figures~\ref{fig:S1}--\ref{fig:S4} present four data sets obtained from three separate devices. All three devices were initially cooled and measured without an applied gate voltage. One of the devices was subsequently thermally cycled, cooled again with a positive gate voltage ($0.3$V) applied, and remeasured; this second cooldown corresponds to the data shown in Fig. \ref{fig:S4}. Each figure includes measurements of the reflection probabilities of quantum point contacts (QPCs), the fractional quasiparticle (QP) charge, and time-domain braiding in both the charged- and neutral-anyon configurations, whose geometries are shown in Figs.~1(a) and 3(a) of the main text, respectively. The corresponding experimental Fano factors are also extracted and compared with the theoretical predictions.
In the charged-anyon configuration, QPC2 serves as the injection QPC and QPC1 as the detection QPC, whereas their roles are reversed in the neutral-anyon configuration, as illustrated in Fig.~\ref{fig:notations}.

The experimental Fano factors were extracted as follows. 
For the configuration targeting charged anyons shown in Fig.~1(a), the experimental charge Fano factor $\mathcal{F}_{\text{charge}}$ was extracted as a function of $e^\ast V_{\rm S}/(2k_{\rm B}T)$ by substituting the measured excess noise $S_{\text{D2}}$ and the tunneling current $I_{\text{QPC1}}$ at the detection QPC into the definition of the charge Fano factor in Eq.~(3) of the main text. 
Similarly, for the neutral-anyon configuration in Fig.~3(a), the experimental neutral Fano factor $\mathcal{F}_{\text{neutral}}$ was extracted by substituting the measured excess noise $S_{\text{D1}}$, the tunneling current $I_{\text{QPC1}}$ at the injection QPC, and the differential reflection probabilities $R_{\text{QPC2}}^{\text{diff}}$ at the detection QPC into Eq.~(4) of the main text. 

The measurement data for time-domain braiding were compared with the theoretical predictions for both the PH-Pf and A-Pf edges. For the charge-anyon configuration shown in Fig.~1(a), the theoretical charge Fano factor $\mathcal{F}_{\text{charge}}^\text{th}$ was obtained by computing the tunneling current $I_{\text{QPC1}}$ and the excess noise $S_{\text{D2}}$ at the detection QPC as a function of the bias voltage $V_{\textrm{S}}$, which took into account the time-domain braiding, the trivial partitioning, and the intermediate processes (Methods and \ref{sec_Fano_total}). For the comparison, their quantitative values were obtained by substituting the experimentally measured reflection probability $R_{\text{QPC2}} (V_{\textrm{S}})$ at the injection QPC and the parameters of the conformal field theories (CFTs) for the ideal PH-Pf and A-Pf edges (the braiding monodromy and the scaling dimensions shown in Table 1 of Methods) into the derived expression in Eq.~\eqref{eq:ChargeFano with rate}. The resulting theoretical Fano factors are shown in panel (d) of Figs.~\ref{fig:S1}--\ref{fig:S4}.
The theoretical curves in panel (c) of Figs.~\ref{fig:S1}--\ref{fig:S4} are the expected noise $S_\text{D2}=2e^* \mathcal{F}_\text{charge}^\text{th} I_\text{QPC1}$, obtained by combining the measured $I_\text{QPC1} (V_{\textrm{S}})$ and the theoretical Fano factor $\mathcal{F}_\text{charge}^\text{th} (V_{\textrm{S}})$.
 
Similarly, for the neutral-anyon configuration shown in Fig.~3(a), the theoretical neutral Fano factor $\mathcal{F}_{\text{neutral}}^\text{th}$ was obtained by computing the differential reflection probability $R_{\text{QPC2}}^{\text{diff}}$ and the excess noise $S_{\text{D1}}$ at the detection QPC as a function of $V_{\textrm{S}}$.
In the calculation, the contributions from the time-domain braiding and trivial partitioning were included, while that of the intermediate process vanishes (\ref{sec_Fano_total}).
For comparison with the measurement data, their quantitative values were obtained by substituting the measured reflection probability $R_{\text{QPC1}} (V_{\textrm{S}})$ at the injection QPC and the CFT parameters for the ideal PH-Pf and A-Pf edges (the braiding monodromy and the scaling dimensions in Table 1) into the derived expression in Eq.~\eqref{eq:NeutralFano with rate}. The resulting theoretical Fano factors are shown in panel (h) of Figs.~\ref{fig:S1}--\ref{fig:S4}.
The theoretical curves in panel (g) of Figs.~\ref{fig:S1}--\ref{fig:S4} are the expected noise $S_\text{D1}=2e^* \mathcal{F}_\text{neutral}^\text{th} I_\text{QPC1} R_\text{QPC2}^\text{diff}$, obtained by combining the theoretical Fano factor $\mathcal{F}_\text{neutral}^\text{th}$ and the measured values of $I_\text{QPC1}(V_\text{S})$ and $R_\text{QPC2}^\text{diff}(V_\text{S})$.

Taken together, the data from all four measurements and in both configurations agree with the PH-Pf predictions within the experimental uncertainty, without any fitting parameters. Taking Fig.~\ref{fig:S1} as a representative example, this agreement is particularly clear in panels (c) and (d) for the charged-anyon configuration, where the PH-Pf and A-Pf predictions are well separate. In the neutral-anyon configuration, shown in panels (g) and (h), the data also agree well with the PH-Pf prediction, although the A-Pf prediction lies closer to the measured values and is therefore less clearly distinguishable. In both configurations, the experimental Fano factors approach the corresponding PH-Pf theoretical values at large source voltages, $|V_\text{S}|$, consistent with the validity regime $e^* |V_\text{S}| \gg k_\text{B} T$ of our time-domain braiding theory. The remaining data sets, shown in Figs. S2–S4, exhibit the same overall behavior.

For clarity, we specify the data sets used for the representative panels in the main text. Figures~1(b) and 2(b) in the main text were obtained from measurement data set 4 in Fig.~\ref{fig:S4}, whereas Fig.~3(b) was obtained from measurement data set 3 in Fig.~\ref{fig:S3}.

\newpage

\clearpage

\section{Theory of Fano factors} \label{sec_Fano_total}

\subsection{Notation} \label{sec_notation}

\begin{figure}[!htbp]
	\centering
	
	\includegraphics[
	width= 0.8\textwidth,
	height=\textheight,
	keepaspectratio
	]{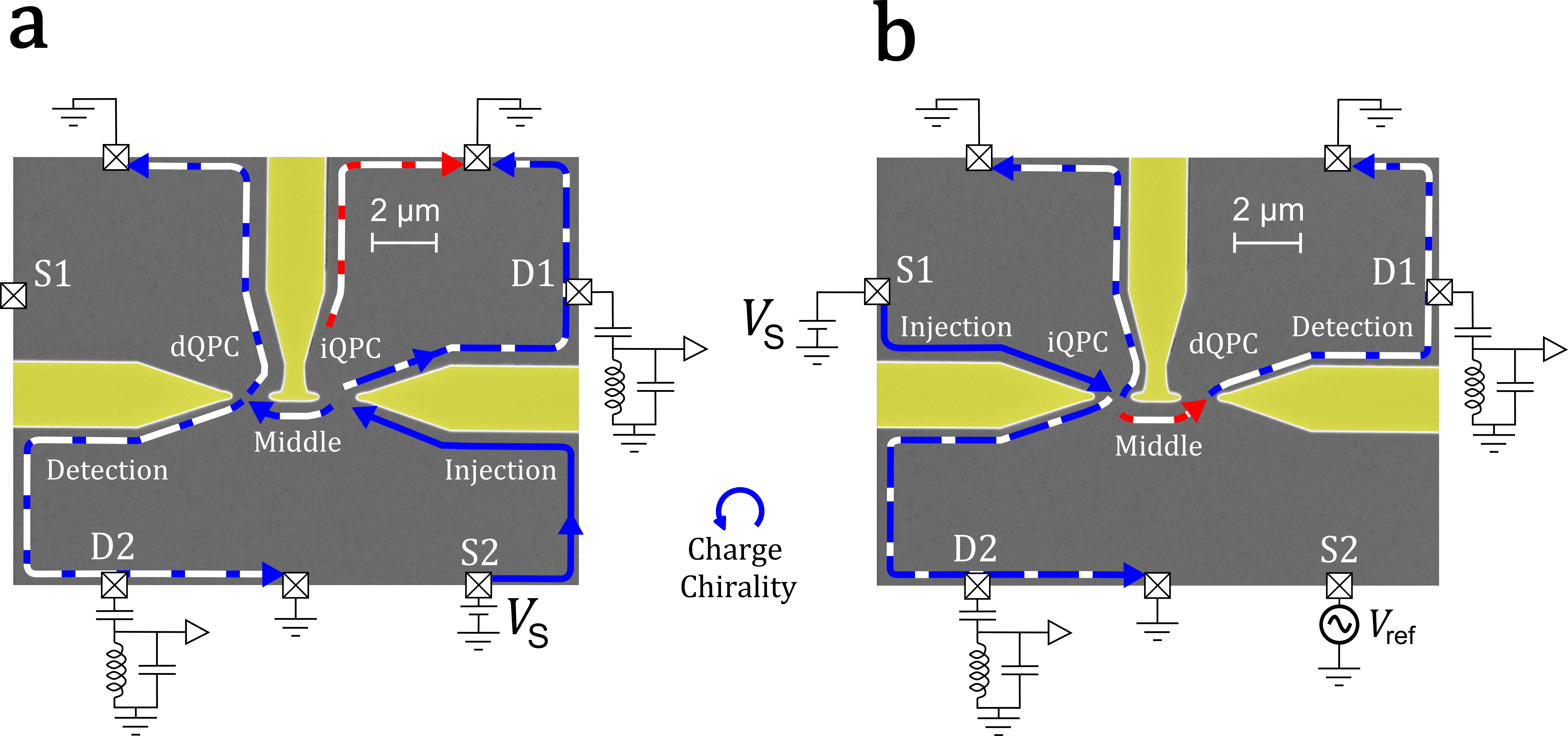}
	\caption{{\bf (a) Experimental configuration} for targeting the braiding of charged anyons. {\bf (b) Experimental configuration} for targeting the braiding of neutral anyons. Throughout the Supplementary Notes, we use role-based notation for QPCs and Edges: the injection QPC (iQPC), the detection QPC (dQPC), and the injection, middle, and detection edges are labeled.}
	\label{fig:notations}
\end{figure}

Our scheme consists of two configurations: the configuration targeting charged anyons, shown in Fig.~1(a) of the main text, and the configuration targeting neutral anyons, shown in Fig.~3(a) of the main text. Throughout the Supplementary Notes, we refer to them as the charged-anyon and neutral-anyon configurations, respectively.

Because the roles of the QPCs and edges differ between these two configurations, we use role-based notation in the Supplementary Notes: the injection QPC, the detection QPC, the injection edge, the middle edge, and the detection edge, as illustrated in Fig.~\ref{fig:notations}. In the charged-anyon configuration [Fig.~\ref{fig:notations}(a)], QPC2 serves as the injection QPC and QPC1 serves as the detection QPC. Correspondingly, Edge3, Edge2, and Edge1 are referred to as the injection, middle, and detection edges, respectively. In the neutral-anyon configuration [Fig.~\ref{fig:notations}(b)], QPC1 serves as the injection QPC and QPC2 serves as the detection QPC, while Edge1, Edge2, and Edge3 are referred to as the injection, middle, and detection edges, respectively.

We sometimes set $\hbar = 1$.

\subsection{Particle-hole Pfaffian (PH-Pf) edge model} \label{sec_PHPf}

\begin{figure}[b]
	\centering
\	\includegraphics[width = .55\textwidth]{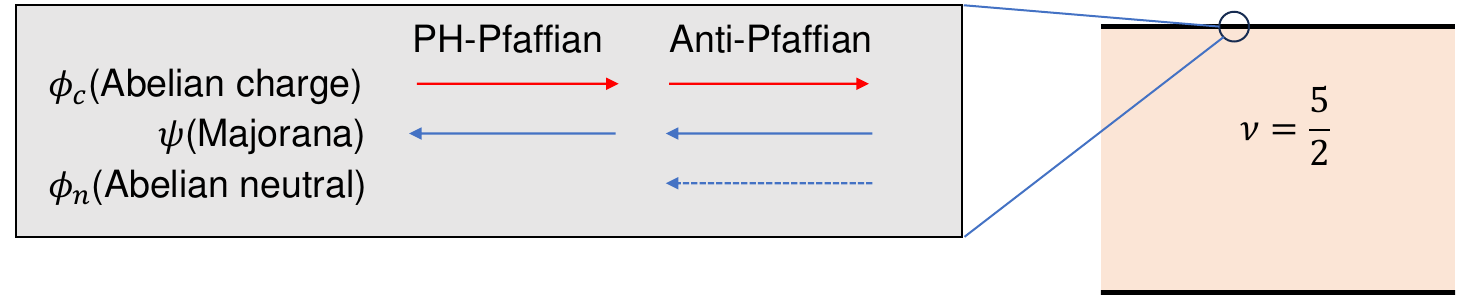}
	\caption{PH-Pf and A-Pf edge models.
	}
	\label{fig:Edge Structure}	
\end{figure}

We provide the model for the PH-Pf edge~\cite{Ma19}. At the fixed point, the $\nu=1/2$ fractional channel of the PH-Pf edge has a downstream charge mode and an upstream Majorana fermion mode (Fig.~\ref{fig:Edge Structure}). It is described by the fixed-point Hamiltonian
\begin{equation} \label{PHPfH}
	H_{0}= \int_{-\infty}^{\infty} dx \Big[ \frac{2 v_{c}}{4\pi}(\partial_{x}\phi_{c}(x))^{2} + \frac{1}{2} v_{n} \psi(x)(i\partial_{x})\psi(x) \Big],
\end{equation} where we omitted the normal ordering symbol. Here $\phi_{c}$ describes the downstream bosonic charge mode with velocity $v_{c}$, and $\psi$ represents the upstream Majorana mode with velocity $v_{n}$. Their commutators are
\begin{equation} \label{PHPfCommutator}
	 [\phi_{c}(x),\phi_{c}(x')]=i\frac{\pi}{2}\textrm{sgn}(x-x')\,, \quad \{\psi(x),\psi(x')\}=\delta(x-x').
\end{equation}

Our experiment probed quasiparticle (QP) tunneling of charge $e^* = e/4$ at the QPCs. The quasiparticle operator must be local with respect to the electron operator $\psi e^{2i\phi_{c}}$.  
For the PH-Pf, there is a single allowed quasiparticle operator carrying the $e^*=e/4$ QP excitation on an edge, 
\begin{equation} \label{PHPfQP}
	\Psi = 
	\frac{1}{\sqrt{2\pi a}} e^{i\phi_{c}/2} \sigma.
\end{equation}
Here $e^{i\phi_{c}/2}$ describes the downstream Abelian charged anyon, $\sigma$ denotes the upstream non-Abelian Ising anyon, and $a$ is the short-length cutoff. The fusion rule of the $\sigma$ anyon is $\sigma \times \sigma = I + \psi$. Tunneling of the QP from position $x_i$ on Edge$i$ to position $x_{j}$ on Edge$j$ across a QPC at time $t$ written as
\begin{equation} \label{PHPfTunneling}
	\mathcal{T}_{\text{QPC}}(t) = \xi_{\text{QPC}} e^{-ie^{*}V t} [\Psi_{j}^{\dagger}(x_{j},t)\Psi_{i} (x_i,t)]_{I}.
\end{equation} 
Here, $\xi_\text{QPC}$ is the tunneling amplitude, $V$ is the voltage difference between the Edges,  
and $[...]_{I}$ means the pair creation of the Ising anyons from the vacuum channel in the tunneling.

\subsection{Anti-Pfaffian (A-Pf) edge model} \label{sec_APf}

We provide the model for the A-Pf edge~\cite{APf1,APf2}. At the fixed point, the A-Pf edge has a  downstream charge mode, an  upstream neutral boson mode, and an upstream Majorana fermion mode (Fig.~\ref{fig:Edge Structure}). Its fixed-point Hamiltonian is
\begin{equation} \label{APfH}
	H_{0}= \int_{-\infty}^{\infty} dx \Big[ \frac{2 v_{c}}{4\pi}(\partial_{x}\phi_{c}(x))^{2} +\frac{v_{n}}{4\pi} (\partial_{x}\phi_{n}(x))^{2} + \frac{1}{2} v_{n} \psi(x)(i\partial_{x})\psi(x) \Big].
\end{equation} 
Here $\phi_{c}$ describes the downstream bosonic charge mode, while $\phi_{n}$ and $\psi$ represent the upstream bosonic neutral mode and the upstream Majorana mode, respectively. The neutral boson mode $\phi_n$ propagates with the same velocity $v_n$ as the Majorana mode due to the underlying symmetry of the SU(2)$_2$ model, and  obeys 
\begin{equation} \label{APfCommutator}
	[ \phi_{n}(x) , \phi_{n}(x') ]=-i\pi\textrm{sgn}(x-x'),
\end{equation}
while the other modes follow Eq.~\eqref{PHPfCommutator}.
The QP operators carrying $e^*=e/4$ on the A-Pf edge must satisfy the locality requirement with respect to the possible electron operators $\psi e^{2i\phi_{c}}$ and $e^{\pm i\phi_{n}}e^{2i\phi_{c}}$.
There are two such possible operators, denoted by $\Psi_{p}$ with $p=\pm 1$,
\begin{equation} \label{APfQP}
	\Psi_{p = \pm 1} = 
	\frac{1}{\sqrt{2\pi a}} e^{i\phi_{c}/2} e^{i p \phi_{n}/2} \sigma.
\end{equation}    
Here $e^{i\phi_{c}/2}$ describes the downstream Abelian charged anyon, and $\sigma e^{i p \phi_{n}/2}$ represents the upstream non-Abelian anyon, consisting of the Abelian sector $e^{i p \phi_{n}/2}$ and the non-Abelian Ising sector $\sigma$. The tunneling operator across a QPC, which annihilates a $p_i = \pm 1$ type of QP excitation at position $x_i$ on Edge$i$ and creates a $p_j = \pm 1$ type of QP excitation at position $x_j$ on Edge$j$, is written as
\begin{equation} \label{APfTunneling}
	\mathcal{T}_{\text{QPC},\,\vec{p} = (p_i, p_j)}(t) = \xi_{\text{QPC},\,\vec{p}} e^{-ie^{*}V t} [\Psi_{j,p_j}^{\dagger}(x_{j},t)\Psi_{i,p_i} (x_i,t)]_{I}\,.    
\end{equation} 
All four tunneling operators $\mathcal{T}_{\text{QPC}, p_i=\pm1, p_j=\pm1}$ are taken into account, as they have the same scaling dimension.  

\subsection{Non-equilibrium correlator} \label{sec_correlator}

The tunneling current and noise at the detection QPC in both configurations were derived by combining the edge models in \ref{sec_PHPf} and \ref{sec_APf} with the Keldysh formalism. The derivation was done in the lowest-order QP tunneling at the detection QPC, while QP tunneling at the injection QPC was treated non-perturbatively (this tunneling was considered up to infinite order). 

To compute the tunneling current and noise at the detection QPC, we considered the non-equilibrium correlator~\cite{Lee22} of QP tunneling operators at time $t_1$ and $t_2$ across the detection QPC in the presence of the source voltage $V_\text{S}$.
In the PH-Pf model, the nonequilibrium correlator is written as
\begin{equation} \label{neqC}
	\mathcal{C}_\text{neq} (t_1, t_2) =
		\Big\langle \mathcal{T}_{\textrm{dQPC}}(0,t_1) \mathcal{T}_{\textrm{dQPC}}^{\dagger} (0,t_2) \Big\rangle_\text{neq} = \frac{|\xi_\text{dQPC}|^2}{d_{\sigma}} G_\text{M,neq}^{>}(t_1-t_2) G_\text{D,eq}^{<}(t_2-t_1)\,,
\end{equation} 
where $\mathcal{T}_{\textrm{dQPC}}$ is the tunneling operator at the detection QPC in Eq.~\eqref{PHPfTunneling}, the QPC location on each edge is set to $x=0$, and $d_{\sigma}=\sqrt{2}$ is the quantum dimension of the QP operator $\Psi$. 
The correlator is decomposed into the Green functions, where the middle edge ($\text{M}$) is described by the non-equilibrium Green function  
$G_\text{M,neq}^> (t_1-t_2) \equiv -i\langle \Psi_{\text{M}} (t_1) \Psi^{\dagger}_{\text{M}}(t_2) \rangle_\text{neq}$, while the detection edge ($\text{D}$) remains in equilibrium, described by $G_\text{D,eq}^< (t_2-t_1) \equiv i\langle \Psi^{\dagger}_{\text{D}} (t_1) \Psi_{\text{D}} (t_2)\rangle_\text{eq}$. 
Here $\langle \cdots \rangle_{\textrm{eq}}$ means the averaging over the thermal states of bosonic fields in equilibrium, $\Psi_{\text{M}}$ and $\Psi_{\text{D}}$ denote QP operators on the middle and detection edges, and the spatial coordinate in the Green functions is omitted. For the A-Pf model, on the other hand,
all possible tunneling operators in Eq.~\eqref{APfTunneling} were taken into account as $\mathcal{C}_\text{neq} (t_1, t_2) = \sum_{\vec{p}_0=(p_{\text{M}0}, p_{\text{D}0})}
\Big\langle \mathcal{T}_{\textrm{dQPC},\vec{p}_0}(0,t_1) \mathcal{T}_{\textrm{dQPC},\vec{p}_0}^{\dagger} (0,t_2) \Big\rangle_{\textrm{neq}}$; there are four possible tunneling operators with $p_{\text{M}0}= \pm 1$ and $p_{\text{D}0}= \pm 1$, as there are two possible QPs [see Eq.~\eqref{APfQP}].
It is decomposed into $\mathcal{C}_\text{neq} (t_1, t_2) \propto \sum_{p_{\text{M}0} = \pm 1, p_{\text{D}0}= \pm 1}  |\xi_{\textrm{dQPC},(p_{\text{M}0},p_{\text{D}0})}|^{2} G_{\text{M,neq;} p_{\text{M}0}}^{>}(t_1-t_2) G_{\text{D,eq;} p_{\text{D}0}}^{<}(t_2-t_1)$.

We now make explicit the connection between the non-equilibrium factor $\Delta(t_1-t_2)$ introduced in Eq.~(7) of the main text and the non-equilibrium Green functions used in the Supplementary Notes. The non-equilibrium Green function factorizes into an equilibrium Green function and a non-equilibrium contribution. Specifically, the PH-Pf edge gives $G_{\text{M,neq}}^{\lessgtr} (t_1-t_2) = G_{\text{M,eq}}^{\lessgtr} (t_1-t_2)\Delta(t_{1}-t_{2})$. For the A-Pf edge, the non-equilibrium factor depends on the quasiparticle types and the corresponding Green function takes the form of $G_{\text{M,neq;} p_{\text{M}0}}^> (t_1-t_2) = G_{\text{M,eq} }^> (t_1-t_2)\Delta_{p_{\textrm{M}0}}(t_{1}-t_{2})$.

The non-equilibrium Green functions such as $G_\text{M,neq}^{>}(t_1-t_2)$ in Eq.~\eqref{neqC} were derived using the Keldysh formalism~\cite{Lee22}. For the PH-Pf edge, the greater Green function on the middle edge is written as 
\begin{equation} \label{Eq:Keldysh_Formalism PHPf}
	G_\text{M,neq}^{>}(t_1-t_2) =\Big\langle T_{K}e^{-i\int_{K} d\tilde{t} H_{T,\text{iQPC}}(\tilde{t}) }(-i)[\Psi_\text{M}(0,t_1^{-})\Psi_\text{M}^{\dagger}(0,t_2^{+})]_{I}\Big\rangle_{\textrm{eq}} =\sum_{m=0}^{\infty} C_{2m}(t_1-t_2).
\end{equation} 
Here, $K$ denotes the Keldysh contour $(-\infty,\infty)\cup (\infty,-\infty)$, the superscripts $\pm$ in the time arguments denote the Keldysh indices, $T_{K}\{\cdots\}$ means operators are contour ordered,  and
$H_{T,\text{iQPC}} = \mathcal{T}_\text{iQPC} + \mathcal{T}_\text{iQPC}^\dagger$ is the tunneling Hamiltonian at the injection QPC. Expanding the Green function in Eq.~\eqref{Eq:Keldysh_Formalism PHPf} in powers of the tunneling Hamiltonian $H_{T,\text{iQPC}}$, the contribution $C_{2m}$ at order $2m$ takes the form,
\begin{equation} \label{c2mPHPf}
	\begin{split}
		C_{2m}(t_1-t_2) = & (-i ) \frac{(-i)^{2m}}{(2m)!}\frac{(2m)!}{ (m!)^2 } \Big[\sum_{\eta_{j}=\pm} \prod_{j=1}^{2m} \eta_{j}\int d\tilde{t}_{j}\Big] \Big\langle T_{K} [\Psi_\text{M}(0,t_1^{-}) \Psi_\text{M}^{\dagger} (0,t_2^{+})]_{I} \prod_{i=1}^{m} \mathcal{T}_{\text{iQPC}}^{\dagger}(\tilde{t}_{2i-1}^{\eta_{2i-1}})\mathcal{T}_{\text{iQPC}}(\tilde{t}_{2i}^{\eta_{2i}}) \Big\rangle_{\textrm{eq}}.
	\end{split}
\end{equation} 
In the order $2m$, each of the tunneling operators $\mathcal{T}^\dagger_{\text{iQPC}}$ and $\mathcal{T}_{\text{iQPC}}$ at the injection QPC appears $m$ times. 

On the other hand, for the A-Pf edge, we consider the two types of $e^*=e/4$ QP operators $\Psi_{e/4,p}$ with $p=\pm 1$ in Eq.~\eqref{APfQP}.  
The greater Green function on the middle edge is written as 
\begin{equation} \label{Eq:Keldysh_Formalism APf}
	G_{\text{M,neq;} p_{\text{M}0}}^{>}(t_1-t_2) = \Big\langle T_{K}e^{-i\int_{K}d\tilde{t} H_{T,\text{iQPC}}(\tilde{t}) }(-i) [\Psi_{\text{M},p_{\text{M}0}}(0,t_1^{-})\Psi_{\text{M},p_{\text{M}0}}^{\dagger}(0,t_2^{+})]_{I} \Big\rangle_{\textrm{eq}} = \sum_{m=0}^{\infty} C^{(p_{\text{M}0})}_{2m}(t_1-t_2),
\end{equation} 
where $\Psi_{\text{M},p_{\text{M}0}}$ denotes the QP operator of $p_{\text{M}0} = \pm 1$ on the middle edge. The tunneling Hamiltonian at the injection QPC takes the form 
$H_{T,\text{iQPC}} = \sum_{\vec{p}=(p_\text{M},p_\text{I}); p_\text{M},p_\text{I}=\pm1} \mathcal{T}_{\text{iQPC}, \vec{p}} + \mathcal{T}_{\text{iQPC}, \vec{p}}^\dagger$.
The perturbative expansion of the Green function in Eq.~\eqref{Eq:Keldysh_Formalism APf} at order $2m$ takes the form
\begin{equation}
\label{c2mAPf}
\begin{aligned}
C^{(p_{\mathrm{M}0})}_{2m}(t_1-t_2)
=& -i \sum_{\{\vec p_i\},\{\vec p'_i\}}
\frac{(-i)^{2m}}{(2m)!}
\frac{(2m)!}
{\displaystyle\prod_{\vec p\in\mathcal P} m_{\vec p}!\,m'_{\vec p}!}
\bigg[
\sum_{\eta_j=\pm}
\prod_{j=1}^{2m}\eta_j
\int d\tilde t_j
\bigg]
\\
&\times
\bigg\langle
T_K
\left[
\Psi_{\mathrm{M},p_{\mathrm{M}0}}(0,t_1^-)
\Psi_{\mathrm{M},p_{\mathrm{M}0}}^\dagger(0,t_2^+)
\right]_I
\prod_{i=1}^{m}
\mathcal T_{\mathrm{iQPC},\vec p_i}^{\dagger}
\!\left(\tilde t_{2i-1}^{\eta_{2i-1}}\right)
\mathcal T_{\mathrm{iQPC},\vec p'_i}
\!\left(\tilde t_{2i}^{\eta_{2i}}\right)
\bigg\rangle_{\mathrm{eq}} .
\end{aligned}
\end{equation}
Here, $\mathcal P=\{(1,1),(1,-1),(-1,1),(-1,-1)\}$,  and \(\vec p_i=(p_{\mathrm{M}i},p_{\mathrm{I}i})\), \(\vec p'_i=(p'_{\mathrm{M}i},p'_{\mathrm{I}i})\), with \(i=1,\ldots,m\). For each \(\vec p\in\mathcal P\), \(m_{\vec p}\) denotes the number of indices \(i\) for which \(\vec p_i=\vec p\), while \(m'_{\vec p}\) denotes the number of indices \(i\) for which \(\vec p'_i=\vec p\). These integers satisfy $\sum_{\vec p\in\mathcal P}m_{\vec p}=m$ and $\sum_{\vec p\in\mathcal P}m'_{\vec p}=m$. Among all possible configurations \(\{\vec p_i\}\) and
\(\{\vec p'_i\}\), only those satisfying the neutrality condition
$
\sum_{i=1}^{m}\vec p_i
=
\sum_{i=1}^{m}\vec p'_i
$
give a nonzero contribution to
\(C^{(p_{\mathrm{M}0})}_{2m}\).

We computed the non-equilibrium correlator $\mathcal{C}_\text{neq} (t_1, t_2)$ [Eq.~\eqref{neqC}] for both PH-Pf and A-Pf edges in the regime of $e^* V_\text{S} \gg k_\text{B} T$.
In this regime, the time-domain braiding process provides the dominant contribution $\mathcal{C}_\text{neq}^\text{TDB} (t_1, t_2)$ to the correlator. Subdominant contributions arise from trivial partitioning and intermediate processes, denoted by $\mathcal{C}_\text{neq}^\text{triv}$ and $\mathcal{C}_\text{neq}^\text{interm}$, respectively. 
Thus, the non-equilibrium correlator can be written as
$\mathcal{C}_\text{neq} (t_1, t_2) \simeq \mathcal{C}_\text{neq}^\text{TDB} (t_1, t_2) + \mathcal{C}_\text{neq}^\text{triv} (t_1, t_2) + \mathcal{C}_\text{neq}^\text{interm} (t_1, t_2) + \cdots$. Below, we present the contribution of each process to the correlator.

\subsubsection{Contribution from time-domain braiding} \label{sec_TDB}

Following the procedures of Refs.~\cite{Lee22,Lee20}, we obtained
the contribution $C_{2m}^{\textrm{TDB}}(t_1-t_2)$ of the time-domain braiding to the $C_{2m}$ term in Eq.~\eqref{Eq:Keldysh_Formalism PHPf} at finite temperature. In Eqs.~\eqref{c2mPHPf} and \eqref{c2mAPf}, time-domain braiding arises from time configurations in which the anyons from $\mathcal{T}^\dagger_{\text{iQPC}}(\tilde{t}_{2i-1})$ and those from $\mathcal{T}_{\text{iQPC}}(\tilde{t}_{2i'})$ overlap pairwise on the middle edge within a short temporal separation of order $\hbar / (e^* V_\text{S})$. Since $e^* V_\text{S}$ is the largest energy, it is a good approximation to set $\tilde{t}_{2i-1} \simeq \tilde{t}_{2i'}$ for each pair. Each such pair represents an anyon in the diluted beam generated at the injection QPC and propagating toward the detection QPC.
Then, in the charged-anyon configuration on the PH-Pf edge, the contribution was found as 
\begin{equation} \label{TDBcharge}
	\begin{split}
		C_{2m}^{\textrm{TDB}}(t_1-t_2) = & G_\text{M,eq}^{>}(t_1-t_2) \frac{(-1)^{m} }{m!}   \Big((1-M_{c}) w_{\text{i} \rightarrow \text{m}} + (1-M_{c}^{*}) w_{\text{m} \rightarrow \text{i}} \color{black}\Big)^{m} (t_1 - t_2)^m \\
		=& G_\text{M,eq}^{>}(t_1-t_2) \frac{(-1)^{m}}{m!} (\frac{I_\text{iQPC}}{e^*})^m \Big( \coth\Big(\frac{e^{*}V_\text{S}}{2 k_\text{B} T}\Big)\textrm{Re}[1-M_{c}] - i \textrm{Im}[M_{c}] \Big)^{m} (t_1 - t_2)^m
	\end{split}
\end{equation} 
for $t_1 - t_2 >0$, while $C_{2m}^{\textrm{TDB}}(t_1-t_2)$ has the same expression but with the replacement of $(t_{1}-t_{2})\rightarrow (t_{2}-t_{1})$, $M_c \to M_c^*$ and $M_c^* \to M_c$ for $t_1 - t_2 < 0$. Here, $G_\text{M,eq}^{>}(t_1-t_2)$ is the equilibrium Green function of the middle edge, $M_c$ denotes the unit braiding monodromy of the charge sector of the QP excitation \eqref{PHPfQP}, $w_{\text{i} \rightarrow \text{m}}$ is the tunneling rate of a single QP from the injection edge (I) to the middle edge (M), and $w_{\text{m} \rightarrow \text{i}}$ is the rate of QP tunneling in the reverse direction. In the second equality of Eq.~\eqref{TDBcharge}, the following relations were used: the current $I_\text{iQPC}$ across the injection QPC by the source voltage $V_\text{S}$ is written as $I_\text{iQPC} = e^* (w_{\text{i} \rightarrow \text{m}} - w_{\text{m} \rightarrow \text{i}} )$, the associated noise is written as $S = 2 (e^*)^2 (w_{\text{i} \rightarrow \text{m}} + w_{\text{m} \rightarrow \text{i}} )$, and the current and the noise are related via $S = 2 e^* I_\text{iQPC} \coth (e^* V_\text{S} / (2 k_\text{B} T))$ at temperature $T$. 
Collecting all contributions from all orders into 
Eq.~\eqref{Eq:Keldysh_Formalism PHPf}, we obtained the time-domain braiding contribution to the greater Green function,  
\begin{align} \label{eq:TDB_correlator}
	\begin{split}
	G_\text{M,neq}^{>, \textrm{TDB}}(t_1-t_2) = & \, G_\text{M,eq}^{>}(t_1-t_2) \
    \Delta_{\textrm{TDB}}^{\textrm{c}} ( t_{1}-t_{2} ) \,, \\
	\Delta_{\textrm{TDB}}^{\textrm{c}} ( t_{1}-t_{2} )  = & \, \exp\Big[-\frac{I_\text{iQPC}}{e^{*}}\coth\Big(\frac{e^{*}V}{2k_\text{B} T}\Big)\textrm{Re}[1-M_{c}]|t_1-t_2| + i\frac{I_\text{iQPC}}{e^{*}} \textrm{Im}[M_{c}](t_1-t_2) \Big].
	\end{split}
\end{align} 
Similarly, in the neutral-anyon configuration on the PH-Pf edge, the contribution was found to have the same form as Eq.~\eqref{eq:TDB_correlator}, but with the replacement of $M_c$ by $M_n$,
\begin{align} \label{eq:TDB_correlator neutral2}
	\begin{split}
	G_\text{M,neq}^{>, \textrm{TDB}}(t_1-t_2) & = \, G_\text{M,eq}^{>}(t_1-t_2) \Delta_{\textrm{TDB}}^{\textrm{n}} (t_1 - t_2) \,, \\ 
    \Delta_{\textrm{TDB}}^{\textrm{n}} (t_1 - t_2)  & = \, \exp\Big[-\frac{I_\text{iQPC}}{e^{*}}\coth\Big(\frac{e^{*}V}{2 k_\text{B} T}\Big)\textrm{Re}[1-M_{n}]|t_1-t_2| + i\frac{I_\text{iQPC}}{e^{*}} \textrm{Im}[M_{n}](t_1-t_2) \Big].
	\end{split}
\end{align} 

For the charged- and neutral-anyon configurations along the A-Pf edge, the contributions of the time-domain braiding to the Green function $G_{\text{M,neq;} p_{\text{M}0}}^{>}(t_1-t_2)$ in Eq.~\eqref{Eq:Keldysh_Formalism APf} follow the same expressions as 
Eqs.~\eqref{eq:TDB_correlator} and \eqref{eq:TDB_correlator neutral2}, respectively. This is because the anyons of the A-Pf edge have the same braiding monodromies
$M_c$ and $M_n$ as those of the PH-Pf edge. 
Although the anyons of the PH-Pf and A-Pf edges have the same braiding monodromy factors, their quantitative contributions differ because the equilibrium Green function 
$G_\text{M,eq}^{>}(t_1-t_2)$ depends on the total scaling dimension $\delta$ of the tunneling QPs, which is $1/4$ for the PH-Pf edge and $1/2$ for the A-Pf edge.

The above expressions apply to the limit where the injected diluted beam is described by Poisson statistics. For relatively less diluted beams, we applied the approximation that tunneling events follow binomial statistics. To incorporate this at finite temperature, we
took into account both particle and hole processes. We denote by $N_{p}$ and  $N_{h}$ the total number of particles and holes incident on the detection QPC during the time interval \(|t_1-t_2|\). 
The net number of particles passing through the QPC is fixed by the source current, $
N_p-N_h=\frac{I_{\rm S}}{e^\ast}|t_1-t_2|$,
while their ratio is determined by the Boltzmann factor,
$
N_p/N_h=\exp[e^\ast V_{\rm S}/(k_\text{B}T)].
$
From these relations, we obtained~\cite{Lee23}
\begin{align}
\label{eq:particlesholesnumber}
N_{p} = \frac{I_S |t_1 - t_2|}{e^*}\frac{e^{e^* V_\text{S} /k_\text{B} T}}{e^{e^* V_\text{S} /k_\text{B} T} -1}, \quad \quad N_{h} = \frac{I_S |t_1 - t_2|}{e^*}\frac{1}{e^{e^* V_\text{S} /k_\text{B} T} -1}\,. 
\end{align}
Using Eq.~\eqref{eq:particlesholesnumber}, the finite-temperature binomial factors replacing the Poisson factors $\Delta_{\textrm{TDB}}^{\textrm{c}}$ and $\Delta_{\textrm{TDB}}^{\textrm{n}}$ in Eqs.~\eqref{eq:TDB_correlator} and \eqref{eq:TDB_correlator neutral2} are given by 
\begin{subequations}
\begin{align} 
\label{eq:exponentialcharge}
	\Delta_{\textrm{TDB}}^{\textrm{c}} (t_{1}-t_{2}) &= \exp(N_p \mathcal{L}_c + N_h \mathcal{L}_c^*) =  \exp\Big[ \frac{I_\text{S} |t_1 - t_2|}{e^*} \Big( \frac{e^{e^* V_\text{S} /k_\text{B} T}}{e^{e^* V_\text{S} /k_\text{B} T} -1} \mathcal{L}_c+ \frac{1}{e^{e^* V_\text{S} /k_\text{B} T} -1} \mathcal{L}_c^*\Big)\Big]\,,  \\ 
   \Delta_{\textrm{TDB}}^{\textrm{n}} (t_{1}-t_{2}) &= \exp(N_p \mathcal{L}_n + N_h \mathcal{L}_n^*)  = \exp\Big[ \frac{I_\text{S} |t_1 - t_2|}{e^*} \Big( \frac{e^{e^* V_\text{S} /k_\text{B} T}}{e^{e^* V_\text{S} /k_\text{B} T} -1} \mathcal{L}_n+ \frac{1}{e^{e^* V_\text{S} /k_\text{B} T} -1} \mathcal{L}_n^*\Big)\Big]\,.
\label{eq:exponentialneutral}
\end{align}
\end{subequations}
Here $I_\text{S} = \frac{e^2 V_\text{S}}{2 h}$
denotes the impinging current on the injection QPC, and $\mathcal{L}_{c}$ and $\mathcal{L}_{n}$ are the binomial factors entering the exponential factor for the charged- and neutral-anyon configurations, respectively, 
\begin{align} \nonumber
	\mathcal{L}_{c/n} =  \log \big [1-R_{\text{iQPC}} + R_{\text{iQPC}} (\textrm{Re} [M_{c/n}] + i\, \text{sgn}(t_1 - t_2) \textrm{Im} [M_{c/n}]) \big ] \,,
\end{align}
where $ R_{\textrm{iQPC}} $ is the reflection probability at the injection QPC. 
In the high dilution limit of $R_{\rm iQPC}\to 0$, these binomial expressions reduce to the Poisson results in Eqs.~\eqref{eq:TDB_correlator} and \eqref{eq:TDB_correlator neutral2}.

\subsubsection{Contribution from trivial partitioning} \label{sec_triv}

\begin{figure}[t]
	\centering
	\includegraphics[width = .63\textwidth]{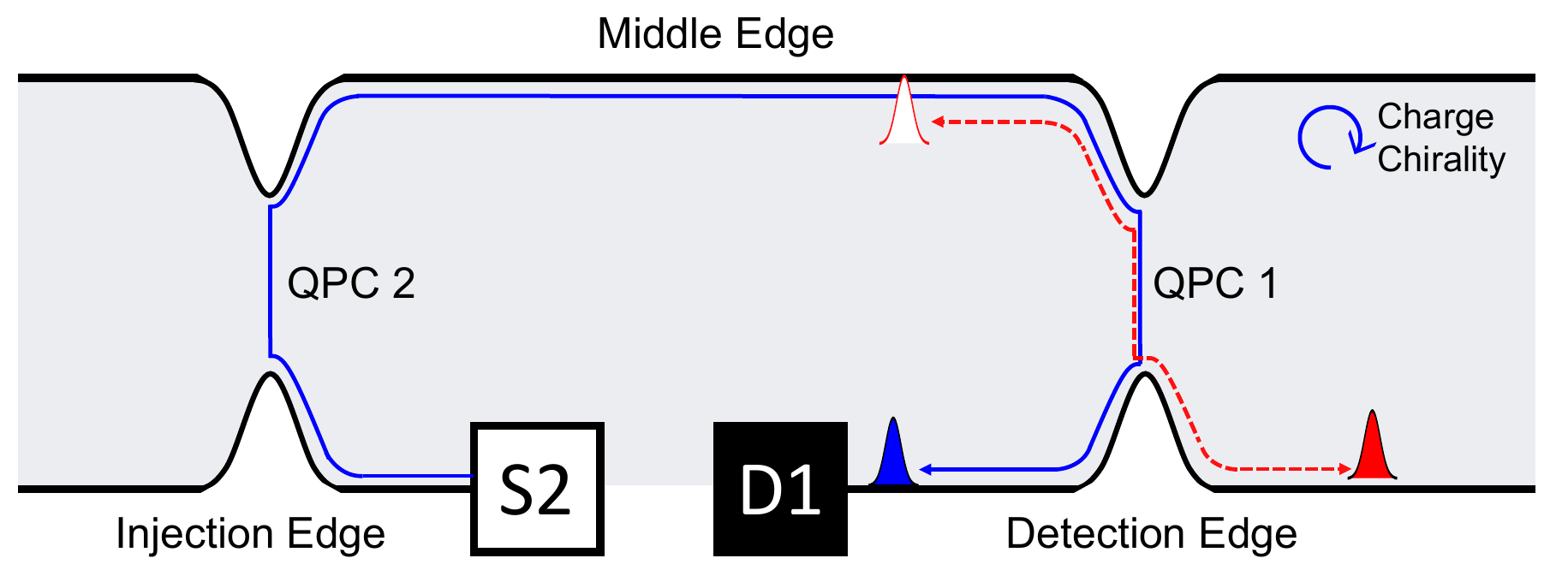}
	\caption{Trivial partitioning of a diluted charged anyon beam in the configuration of Fig.~1(a) of the main text. Tunneling of QPs across QPC2 (injection QPC), activated by the source voltage $V_\text{S}$ applied to S2, generates the diluted charged anyon beam, which propagates toward QPC1 (detection QPC) along the middle edge. The trivial partitioning happens such that a charged anyon (narrow blue wavepacket) from the diluted beam undergoes tunneling from the middle edge to the detection edge upon its arrival at the detection QPC. In the case of the PH-Pf edge, the charged anyon tunnels together with a neutral Ising anyon (narrow red wavepacket), forming a QP. The tunneling leaves another neutral Ising anyon (narrow red-outlined empty wavepacket) on the middle edge. The two Ising anyons are pair-created from the vacuum channel. In the case of the A-Pf edge, the tunneling of the charged anyon is accompanied by either a particle-like neutral anyon (narrow red wavepacket) or a hole-like neutral anyon (not shown). This QP tunneling leaves a hole-like neutral anyon (narrow red-outlined empty wavepacket) or a particle-like neutral anyon (not shown), respectively, on the middle edge. The non-Abelian sectors of the two neutral anyons are pair-created from the vacuum channel.
	In both PH-Pf and A-Pf edges, the QP tunneling involves the energy $e^* V_\text{S}$. 
	}
	\label{fig:trivial partitioning}	
\end{figure}

The trivial partitioning is illustrated in Fig.~\ref{fig:trivial partitioning}.
In this partitioning, tunneling occurs at the detection QPC when an anyon from the diluted beam arrives there. The relevant energy scale of this process is $e^\ast V_\text{S}$, since the tunneling event is directly induced by the non-equilibrium anyon arriving at the detection QPC. 
In the regime of $R_\text{iQPC} \ll 1$ at zero temperature,
the contribution of this process to the tunneling current and noise across the detection QPC is sub-dominant compared to that of the braiding process,
\begin{equation} \label{triv vs TDB}
	\frac{\text{contribution of the trivial partitioning}}{\text{contribution of the time-domain braiding}} \propto R_\text{iQPC}^{2-2\delta},
\end{equation}
similarly to the case of Abelian anyons at $\nu = 1/3$ \cite{Lee23}, as will be shown in \ref{sec_Fano}.

The trivial partitioning process contributes to the second-order perturbation term $C_{2}$ in Eq.~\eqref{Eq:Keldysh_Formalism PHPf}. For the charged-anyon configuration on the PH-Pf edge, $C_{2}$ is written as
\begin{equation} \nonumber
	\begin{split}
		C_2 (t_1 - t_2) = & (-i) (-i)^2 |\xi_\text{iQPC}|^2 \sum_{\eta_{1}\eta_{2}} \eta_{1}\eta_{2} \int d\tilde{t}_{1}d\tilde{t}_{2} 
		e^{ie^{*}V_\text{S} (\tilde{t}_{1}-\tilde{t}_{2})} \\
		\times & \big\langle T_{K} [\Psi_\text{M}(0,t_1^-) \Psi_\text{M}^{\dagger}(0,t_2^+)]_{I} [\Psi_\text{I}^{\dagger}(0,\tilde{t}_{1}^{\eta_{1}})\Psi_\text{M}(-d,\tilde{t}_{1}^{\eta_{1}})]_{I}
		[\Psi_\text{M}^\dagger (-d,\tilde{t}_{2}^{\eta_{2}})\Psi_\text{I} (0,\tilde{t}_{2}^{\eta_{2}})]_{I} \rangle_\text{eq},
	\end{split}
\end{equation}  
where the spatial coordinate of the injection QPC is $x= -d$ on the middle edge and $x=0$ on the injection edge.
The trivial partitioning of a charged anyon from the diluted beam happens upon its arrival at the detection QPC. This corresponds to the time window of 
\begin{equation} \nonumber
	\tilde{t}_2+\frac{d}{v_{c}}\simeq t_{1}, \quad \quad \tilde{t}_1 +\frac{d}{v_{c}}\simeq t_{2}.
\end{equation} 
Applying this condition to the above expression of $C_2$, we obtained the contribution $C_{2}^{\textrm{triv}}(t_1-t_2)$ of the trivial process to the term $C_2$,
\begin{equation} \nonumber
	\begin{split}
		C_2^\text{triv} (t_1-t_2) = & (-i)^2 G_{\textrm{M,eq}}^{>}(t_1-t_2) \frac{|\xi_\text{iQPC}|^2}{4\pi^{2}a^{2}}\frac{1}{d_{\sigma}} \sum_{\eta_{1}\eta_{2}} \eta_{1}\eta_{2} \int d\tilde{t}_{1}d\tilde{t}_{2} e^{ie^{*}V_\text{S}(\tilde{t}_{1}-\tilde{t}_{2}) } \Big( \frac{\pi k_{\textrm{B}} T av_{n}^{{-1}}}{\sin[\pi k_{\textrm{B}} T \{a v_{n}^{{-1}} + i\chi_{\eta_{1}\eta_{2}}(\tilde{t}_{1}-\tilde{t}_{2})(\tilde{t}_{1}-\tilde{t}_{2}) \}]} \Big)^{2\delta_{n}}  \\
		\times &  \Big(\frac{\pi k_{\textrm{B}} T a v_{c}^{-1}}{\sin[\pi k_{\textrm{B}} T \{a v_{c}^{-1} -i\eta_{1} (\tilde{t}_{1}-t_2 + d /v_c)\}]}\Big)^{\delta_{c}} \Big(\frac{\pi k_{\textrm{B}} T a v_{c}^{-1}}{\sin[\pi k_{\textrm{B}} T \{a v_{c}^{-1} -i\eta_{2}(\tilde{t}_{2}-t_1 + d /v_c)\}]}\Big)^{\delta_{c}} \\
		= & (-i)^2 G_{\textrm{M,eq}}^{>}(t_1-t_2) \frac{|\xi_\text{iQPC}|^2}{4\pi^{2}a^{2}}\frac{1}{d_{\sigma}} \sum_{\eta_{1}\eta_{2}} \eta_{1}\eta_{2} \int d\tilde{t}_{1} d\tilde{t}_{2} F^{\eta_1 \eta_2}_{n}(\tilde{t}_{1}-\tilde{t}_{2}) F^{\eta_1 \eta_2}_{c}(\tilde{t}_{1},\tilde{t}_{2},t_1,t_2),
	\end{split}
\end{equation}
where $v_{c(n)}$ is the velocity of the charged (neutral) anyon, $\chi_{\eta_1 \eta_2} (\tilde{t}_1 - \tilde{t}_2) = [(\eta_1 + \eta_2) \text{sgn} (\tilde{t}_1 - \tilde{t}_2) - (\eta_1 - \eta_2)]/2$, and $a$ is the short length cutoff. The integrand has two functions, as each QP tunneling at the injection QPC generates a charged anyon moving to the detection QPC along the middle edge
and a neutral anyon moving in the opposite direction and hence not participating in the trivial partitioning process.
The first function $F^{\eta_1\eta_2}_{n}(\tilde{t}_{1}-\tilde{t}_{2}) = \Big( \frac{\pi k_{\textrm{B}} T av_{n}^{{-1}}}{\sin[\pi k_{\textrm{B}} T \{a + i\chi_{\eta_{1}\eta_{2}}(\tilde{t}_{1}-\tilde{t}_{2})(\tilde{t}_{1}-\tilde{t}_{2}) \}]} \Big)^{2\delta_{n}}$ describes the neutral anyon irrelevant to the trivial partitioning, while the second function 
$F^{\eta_1\eta_2}_{c}(\tilde{t}_{1},\tilde{t}_{2},t_1,t_2) = e^{ie^{*}V_\text{S} (\tilde{t}_{1} - \tilde{t}_{2})} \Big(\frac{\pi k_{\textrm{B}} T a v_{c}^{-1}}{\sin[\pi k_{\textrm{B}} T \{a v_{c}^{-1} -i\eta_{1} (\tilde{t}_{1}-t_2+ d /v_c)\}]}\Big)^{\delta_{c}} \Big(\frac{\pi k_{\textrm{B}} T a v_{c}^{-1}}{\sin[\pi k_{\textrm{B}} T \{a v_{c}^{-1} -i\eta_{2}(\tilde{t}_{2}-t_1+ d /v_c)\}]}\Big)^{\delta_{c}}$ describes the charged anyon which participates in the trivial partitioning process.
In the regime of $e^* V_\text{S} / (k_\text{B}T) \gg1$, the integral of $\int d\tilde{t}_{1} d\tilde{t}_{2} F^{\eta_1 \eta_2}_{n} F^{\eta_1 \eta_2}_{c}$ with $(\eta_1, \eta_2)=(-1,1)$ dominates over the integrals with the other $(\eta_1, \eta_2)$'s, which are smaller by the factor of $e^{- e^* V_\text{S} / (k_\text{B}T)}$. Around the time window of $\tilde{t}_{1}+\frac{d}{v_{c}} - t_2 \in [ - \frac{1}{e^{*}V_\text{S}}, \frac{1}{e^{*}V_\text{S}}]$ and $\tilde{t}_{2}+\frac{d}{v_{c}} -t_1 \in [ - \frac{1}{e^{*}V_\text{S}}, \frac{1}{e^{*}V_\text{S}}]$, 
$F^{-1,1}_{c}(\tilde{t}_1, \tilde{t}_2)$ shows a sharp peak structure representing the trivial partitioning. In the zero-temperature limit, the peak occurs at $\tilde{t}_{1,\text{saddle}} = t_2 -\frac{i\delta_{c}}{e^{*}V_\text{S}}$ and $\tilde{t}_{2, \text{saddle}}= t_1 +\frac{i\delta_{c}}{e^{*}V_\text{S}}$, which satisfy $\partial \ln F^{-1,1}_c  / \partial \tilde{t}_1 = 0$
and $\partial \ln F^{-1,1}_c  / \partial \tilde{t}_2 = 0$, respectively.
Since $F^{-1,1}_n(\tilde{t}_1, \tilde{t}_2)$ varies slowly around the peak position, 
we applied the `saddle-point' approximation,
\begin{equation} \label{saddle-pt1}
F^{-1,1}_n(\tilde{t}_1, \tilde{t}_2) \simeq F^{-1,1}_n(\tilde{t}_{1, \text{saddle}}, \tilde{t}_{2, \text{saddle}}), \quad \quad \int d\tilde{t}_{1} d\tilde{t}_{2} F^{-1,1}_{n} F^{-1,1}_{c} \simeq F^{-1,1}_n(\tilde{t}_{1, \text{saddle}}, \tilde{t}_{2, \text{saddle}}) \int d\tilde{t}_{1} d\tilde{t}_{2} F^{-1,1}_{c}.
\end{equation} 
Computing $\int d\tilde{t}_{1} d\tilde{t}_{2} F^{-1,1}_{c}$ and 
combining the result with Eq.~\eqref{Eq:Keldysh_Formalism PHPf}, we obtained the trivial-partitioning contribution to the Green function as 
\begin{equation} \label{Eq:C_Direct}
		\begin{split}
		G_\text{M,neq}^{>, \textrm{triv}}(t_1-t_2)_{\textrm{neq}} = & \,\, G_\text{M,eq}^{>}(t_1-t_2) \,\Delta_{\textrm{triv}}^{\textrm{c},\textrm{PH-Pf}} (t_{1}-t_{2})\\
        \Delta_{\textrm{triv}}^{\textrm{c},\textrm{PH-Pf}} (t_{1}-t_{2}) =  & \,  C_\text{rd}^\text{c,PH-Pf} R_\text{iQPC}   f^\text{c}_{\textrm{triv}}(\delta_c,\delta_n,\frac{e^{*}V_\text{S}}{k_\text{B}T}) e^{-ie^{*}V_\text{S} (t_1-t_2)} \\
        f^\text{c}_{\textrm{triv}} (\delta_c,\delta_n,\alpha \equiv \frac{e^{*}V_\text{S}}{k_\text{B}T}) = & \,\, \frac{2^{-2-2\delta_{n}}}{\pi^{2}} \Big(\frac{\alpha}{2}\Big) \nu \Big(\frac{e}{e^{*}}\Big)^{2} e^{|\alpha|}\csch\big(\frac{\alpha}{2}\big) \Big(\sin\Big[ \frac{2 \pi \delta_{c} }{|\alpha|}\Big]\Big)^{-2\delta_{n}}  
		\frac{B^{2}(\frac{\delta_{c}}{2} - i\frac{\alpha}{2\pi} , \frac{\delta_c}{2} + i\frac{\alpha}{2\pi})}{B(\delta - i\frac{\alpha}{2\pi} , \delta + i\frac{\alpha}{2\pi})} \\
		\end{split}
\end{equation}
where 
$B(x,y) = \Gamma(x) \Gamma (y)/\Gamma(x+y)$ is the beta function, 
$\Gamma(x)$ is the gamma function, $\nu = 1/2$,
and $\delta = \delta_c + \delta_n$ is the total scaling dimension.
The  reflection probability at the injection QPC  is expressed as
\begin{equation} \label{RiQPC}
	R_\text{iQPC}= |\xi_{\textrm{iQPC}}|^{2}\Big(\frac{2\pi a k_\text{B} T}{v_{c}}\Big)^{2\delta_{c}-1} \Big(\frac{2\pi a k_\text{B} T}{v_{n}}\Big)^{2\delta_{n}} \frac{1}{\nu}(\frac{e^{*}}{e})^{2} \frac{1}{\pi (e^{*} V_\text{S}) a  v_{c} d_{\sigma}} B(\delta+\frac{ie^{*} V_\text{S}}{2\pi k_\text{B} T},\delta-\frac{ie^{*} V_\text{S}}{2\pi k_\text{B} T})\sinh(\frac{e^{*} V_\text{S}}{2 k_\text{B} T})
\end{equation}
for the PH-Pf edge model, while it has the same expression but with the replacement of
$|\xi_{\textrm{iQPC}}|^{2} \rightarrow \sum_{\mathbf{p}}|\xi_{\textrm{iQPC},\mathbf{ p }}|^{2}$ for the A-Pf edge.
It should be noted that the saddle-point approximation in Eq.~\eqref{saddle-pt1} overestimates the integral. In this approximation, the sharply peaked factor $F^{-1,1}_{c}$ is used to identify the saddle-point region, while
$F^{-1,1}_{n}$ is treated as a slowly varying prefactor over that region.
Although $F^{-1,1}_{n}$ varies more slowly than $F^{-1,1}_{c}$, its residual variation over the width of the saddle-point region is not negligible. This leads to an overestimation of the integral.
To compensate for this overestimation, we introduced a reduction factor $C_\text{rd}^\text{c,PH-Pf} \simeq 0.47$, the value of which was determined by directly comparing numerical integration results obtained with and without the `saddle-point' approximation. 
Since the contribution from the trivial partitioning is subdominant, our conclusion regarding the comparison between the experimental data and the theoretical predictions, shown in Figs.~2(b) and 3(b) of the main text and in \ref{sec:comparison},
remains robust against slight variations in the reduction factor or the details of the integration.

We next consider the trivial partitioning in the neutral-anyon configuration on the PH-Pf edge.
In the same manner as for the charged beam, we derived its contribution,
\begin{equation} \label{Eq:C_Direct neutral}
	\begin{split}
		G_\text{M,neq}^{>, \textrm{triv}}(t_1-t_2)_{\textrm{neq}} = & \,\,  G_\text{M,eq}^{>}(t_1-t_2)\, \Delta_{\textrm{triv}}^{\textrm{n},\textrm{PH-Pf}} (t_{1}-t_{2}), \\
        \Delta_{\textrm{triv}}^{\textrm{n},\textrm{PH-Pf}} (t_{1}-t_{2}) =  & \,\,  C_\text{rd}^\text{n,PH-Pf} R_\text{iQPC}   f^\text{n}_{\textrm{triv}}(\delta_c,\delta_n,\frac{e^{*}V_\text{S}}{k_\text{B}T},d_{\sigma}) \times 2 \cos [e^* V_\text{s} (t_1 - t_2)], \\
		f^\text{n}_{\textrm{triv}} (\delta_c,\delta_n,\alpha \equiv \frac{e^{*}V_\text{S}}{k_\text{B}T}, d_{\sigma} ) = & \,\, \frac{2^{-2-2\delta_{c}}}{\pi^{2} d_{\sigma}} \Big(\frac{\alpha}{2}\Big) \nu \Big(\frac{e}{e^{*}}\Big)^{2} e^{|\alpha|}\csch\big(\frac{\alpha}{2}\big) \Big(\sin\Big[ \frac{2 \pi \delta_{n} }{|\alpha|}\Big]\Big)^{-2\delta_{c}}  
		\frac{B^{2}(\frac{\delta_{n}}{2} - i\frac{\alpha}{2\pi} , \frac{\delta_n}{2} + i\frac{\alpha}{2\pi})}{B(\delta - i\frac{\alpha}{2\pi} , \delta + i\frac{\alpha}{2\pi})}, \\
	\end{split}
\end{equation} 
where $C_\text{rd}^\text{n,PH-Pf} \simeq 0.44$.

For the charged-anyon configuration on the A-Pf edge, we found that the contribution of the trivial partitioning to the Green function $G_{\text{M,neq;} p_{\text{M}0}}^{>}(t_1-t_2)$ in Eq.~\eqref{Eq:Keldysh_Formalism APf} follows the same expressions as Eq.~\eqref{Eq:C_Direct}, except that $C_\text{rd}^\text{c,PH-Pf}$ is replaced by $C_\text{rd}^\text{c,A-Pf} \simeq 0.09$. While the PH-Pf and A-Pf cases share the same form, their quantitative contributions differ due to distinct scaling dimensions.

For the neutral-anyon configuration on the A-Pf edge, on the other hand, the trivial-partitioning contribution to the greater Green function for the $p_{\rm M0}$ neutral anyon reads
\begin{equation}
    \begin{split}
        G_{\textrm{M,neq},p_{\textrm{M}_{0}}}^{>,\textrm{triv}}(t_{1}-t_{2}) = & \,\, G_\text{M,eq}^{>}(t_1-t_2)\, \Delta_{\textrm{triv}, p_{\textrm{M}_{0}}}^{\textrm{n}} (t_{1}-t_{2})\\
        \Delta_{\textrm{triv}, p_{\textrm{M}_{0}}}^{\textrm{n}} (t_{1}-t_{2}) =  & \,\,  C_\text{rd}^\text{n,A-Pf} R_\text{iQPC}   f^\text{n}_{\textrm{triv}}(\delta_c,\delta_n,\frac{e^{*}V_\text{S}}{k_\text{B}T},d_{\sigma}) \\
        \times & \Big[\frac{\sum_{p_{\textrm{I}}=\pm 1}|\xi_{\textrm{iQPC},(p_{\textrm{M}_{0}},p_{\textrm{I}})}|^{2}}{\sum_{\mathbf{p}} |\xi_{\textrm{iQPC},\mathbf{p}}|^{2} } e^{-ie^{*} V_\text{S} (t_{1} - t_{2}) } + \frac{\sum_{p_{\textrm{I}}=\pm}|\xi_{\textrm{iQPC},(-p_{\textrm{M}_{0}},p_{\textrm{I}})}|^{2}}{\sum_{\mathbf{p}} |\xi_{\textrm{iQPC},\mathbf{p}}|^{2} } e^{ie^{*} V_\text{S} (t_{1} - t_{2}) } \Big] 
    \end{split}
\end{equation}
with $C_\text{rd}^\text{n,A-Pf} \simeq 0.59$. 
The two terms in the square brackets distinguish the type of neutral anyons arriving at the detection QPC. If the arriving anyon has the same neutral label $p_{\rm M0}$ as the operator in the Green function at the detection QPC, the corresponding contribution carries the phase factor $e^{-i e^\ast V_{\rm S}(t_1-t_2)}$. If instead the incoming anyon has the opposite neutral label $-p_{\rm M0}$, the conjugate process contributes and the phase factor is reversed, $e^{+i e^\ast V_{\rm S}(t_1-t_2)}$. 
The weights of these two processes are determined by the corresponding tunneling amplitudes at the injection QPC. 

\subsubsection{Contribution from intermediate process} \label{sec_interm}

\begin{figure}[t]
	\centering
	\includegraphics[width = .99\textwidth]{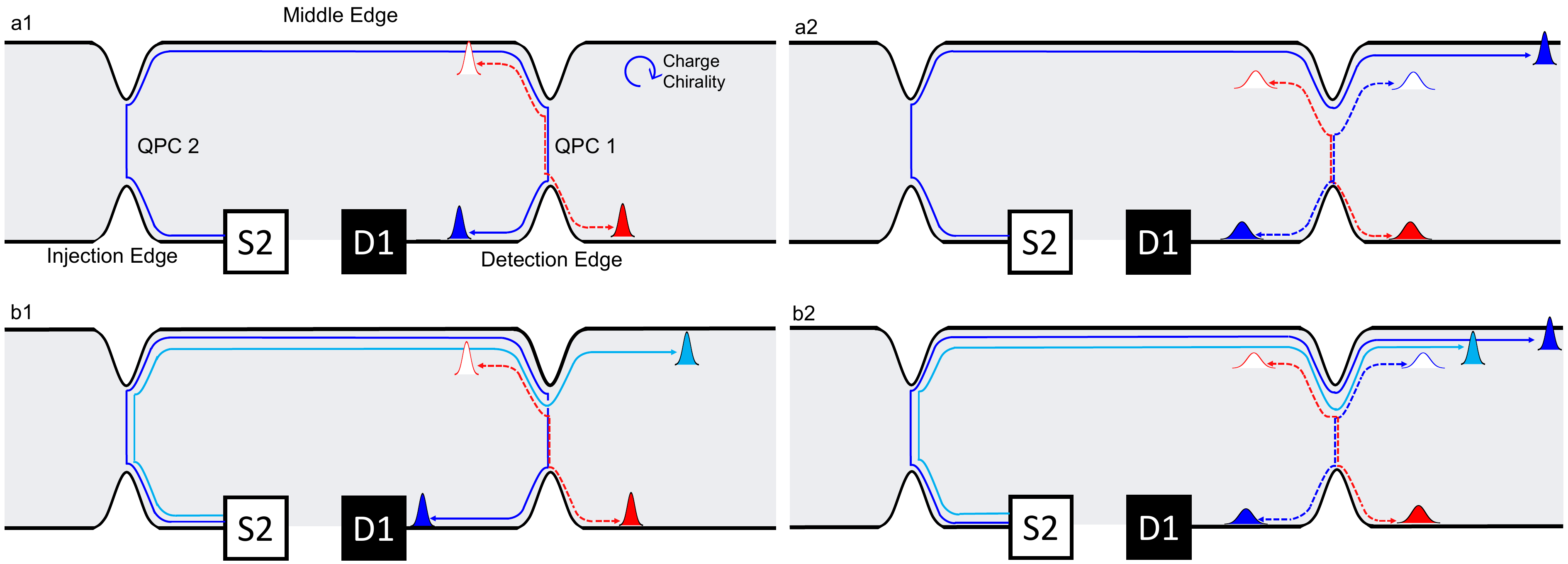}
	\caption{Intermediate processes for a diluted charged-anyon beam in the configuration shown in Fig.~1a of the main text. (a) The lowest-order intermediate process. This process is described by the interference $\text{a}1\text{a}2^*$ between the subprocesses shown in (a1) and (a2). In (a1), a charged anyon (narrow blue wavepacket) from the diluted beam tunnels from the middle edge to the detection edge at the detection QPC (QPC1) at time $t_2$, as in the trivial partitioning of Fig.~\ref{fig:trivial partitioning}. 
This tunneling event is accompanied by the creation of a neutral particle-hole pair (narrow red wavepacket and narrow red-outlined empty wavepacket) at the QPC, whose type depends on whether the edge is PH-Pf or A-Pf. In (a2), the charged anyon passes the detection QPC without tunneling. Subsequently, at time $t_1$, a particle-hole pair of QPs is thermally generated at the QPC, as in the time-domain braiding.
     Wide blue (red) and wide blue(red)-outlined empty wavepackets represent the resulting particle-like and hole-like charged (neutral) anyons. (b) The next-order intermediate process. In addition to the process in (a), another charged anyon from the diluted beam passes the detection QPC between the time $t_2$ of the event in (a1) and the time $t_1$ of the event in (a2). Through the interference $\text{b}1\text{b}2^*$ between the subprocesses shown in (b1) and (b2), this additional anyon is braided by the time-domain loop formed by the interference between (a1) and (a2).
}\label{fig:intermediate}	
\end{figure}

There happens another subdominant process, which we call intermediate process, as its nature is intermediate between the trivial partitioning and the time-domain braiding. 
Figure~\ref{fig:intermediate} illustrates the case of the charged-anyon configuration. 
Its lowest-order tunneling process arises from the interference between two tunneling events at the detection QPC. The first event [see Fig.~\ref{fig:intermediate}(a1)] is a QP tunneling with energy $e^* V_\text{S}$ at time $t_2$, which is directly induced by an anyon from the diluted beam as in the trivial partitioning. The second [see Fig.~\ref{fig:intermediate}(a2)] is a virtual (or thermal) QP tunneling with energy $k_\text{B} T$ at $t_1$, analogous to the QP tunneling at the detection QPC in the time-domain braiding. The interference between the two events forms a time-domain loop at the detection QPC when $|t_1 - t_2| \lesssim \hbar / (k_\text{B} T)$, similarly to the loop in the time-domain braiding;
only one event (a2) thermally occurs in the intermediate process, while both events at $t_1$ and $t_2$ thermally occur in the time-domain braiding.
This time-domain loop braids additional anyons from the diluted beam which pass the detection QPC within the time window between $t_1$ and $t_2$ [Fig.~\ref{fig:intermediate}(b)].
We note that a process corresponding to the lowest-order intermediate process shown in Fig.~\ref{fig:intermediate}(a) has been introduced in Ref.~\cite{Han16}.

In the regime of $R_\text{iQPC} \ll 1$ at zero temperature,
the contribution of the intermediate process to the tunneling current and noise across the detection QPC is sub-dominant compared to that of the time-domain braiding,
\begin{equation} \label{interm vs TDB}
	\frac{\text{contribution of the intermediate process}}{\text{contribution of the time-domain braiding}} \propto  R_\text{iQPC}, 
\end{equation}
which will be shown in \ref{sec_Fano}.

The lowest-order intermediate process occurs in the second-order perturbation term $C_{2}$ in Eq.~\eqref{Eq:Keldysh_Formalism PHPf}. We derive its contribution $C_{2}^{\textrm{interm}}(t_1-t_2)$ to the $C_{2}$ term at finite temperature. 
We first consider the lowest-order intermediate process in the charged-anyon configuration on the PH-Pf edge.
In this case, $C_{2}$ is written as
\begin{equation} \nonumber
	\begin{split}
		C_2 (t_1 - t_2) = & (-i) (-i)^2 |\xi_\text{iQPC}|^2 \sum_{\eta_{1}\eta_{2}} \eta_{1}\eta_{2} \int d\tilde{t}_{1}d\tilde{t}_{2}  
		e^{ie^{*}V_\text{S} (\tilde{t}_{1}-\tilde{t}_{2})} \\
		\times & \big\langle T_{K} [\Psi_\text{M}(0,t_1^-) \Psi_\text{M}^{\dagger}(0,t_2^+)]_{I} [\Psi_\text{I}^{\dagger}(0,\tilde{t}_{1}^{\eta_{1}})\Psi_\text{M}(-d,\tilde{t}_{1}^{\eta_{1}})]_{I}
		[\Psi_\text{M}^\dagger (-d,\tilde{t}_{2}^{\eta_{2}})\Psi_\text{I} (0,\tilde{t}_{2}^{\eta_{2}})]_{I} \rangle_\text{eq}.
	\end{split}
\end{equation}  
The lowest-order intermediate process of a charged anyon from the diluted beam happens in the time windows of 
\begin{equation} \nonumber
\text{(i)} \,\, \tilde{t}_{1}+\frac{d}{v_{c}}\simeq t_2, \quad \quad \tilde{t}_{1}\simeq \tilde{t}_{2},
\quad \quad \quad \quad \quad \text{(ii)} \,\, \tilde{t}_{2}+\frac{d}{v_{c}}\simeq t_1, \quad \quad \tilde{t}_{1}\simeq \tilde{t}_{2}
\end{equation} 
Here, $\tilde{t}_{1}+\frac{d}{v_{c}}\simeq t_2$ in (i) and $\tilde{t}_{2}+\frac{d}{v_{c}}\simeq t_1$ in (ii) correspond to the event in Fig.~\ref{fig:intermediate}(a1), while $\tilde{t}_{1}\simeq \tilde{t}_{2}$ corresponds to Fig.~\ref{fig:intermediate}(a2).
When $t_1 > t_2$, we found that $C_{2}^{\textrm{interm}}(t_1-t_2)$ is nonvanishing under the condition (i).
In this case, the application of the condition (i) to the expression of $C_2$ leads to
\begin{equation} \nonumber
	\begin{split}
	C_{2}^{\textrm{interm}}(t_1-t_2) = & (-i)^2 G_{\textrm{M,eq}}^{>}(t_1-t_2) \frac{|\xi_\text{iQPC}|^2}{4\pi^{2}a^{2} d_{\sigma}}
	\Big[\sum_{\eta_{1}\eta_{2}}\eta_{1}\eta_{2} \int d\tilde{t}_{1} d\tilde{t}_{2} e^{ie^{*}V_\text{S}(\tilde{t}_{1}-\tilde{t}_{2}) }
	 \Big( \frac{\pi k_{\textrm{B}} T av_{c}^{{-1}}}{\sin[\pi k_{\textrm{B}} T \{a v_{c}^{{-1}} + i\chi_{\eta_{1}\eta_{2}}(\tilde{t}_{1}-\tilde{t}_{2})(\tilde{t}_{1}-\tilde{t}_{2}) \}]} \Big)^{2\delta_{c}} \\
	 \times & \Big( \frac{\pi k_{\textrm{B}} T av_{n}^{{-1}}}{\sin[\pi k_{\textrm{B}}T \{a v_{n}^{{-1}} + i\chi_{\eta_{1}\eta_{2}}(\tilde{t}_{1}-\tilde{t}_{2})(\tilde{t}_{1}-\tilde{t}_{2}) \}]} \Big)^{2\delta_{n}} 
	 \frac{(\sin[\pi k_{\textrm{B}} T\{av_{c}^{-1} -i\eta_{2}(\tilde{t}_{2} - t_2 +\frac{d}{v_{c}})\}])^{\delta_{c}} }{(\sin[\pi k_{\textrm{B}} T\{av_{c}^{-1} - i\eta_{1} (\tilde{t}_{1} - t_2 +\frac{d}{v_{c}})\}])^{\delta_{c}}}e^{\frac{i\pi\delta_{c}}{2}(\eta_{1}-\eta_{2})}\Big].
	\end{split}
\end{equation}
For further computation, we used the approximation of $\sin[\pi T\{a-i\eta_{1(2)}(\tilde{t}_{1(2)} - t_2 +\frac{d}{v_{c}})\}]\simeq \pi T\{a-i\eta_{1(2)}(\tilde{t}_{1(2)} - t_2 +\frac{d}{v_{c}})\}$ because the region of $\tilde{t}_{1(2)} - t_2 +\frac{d}{v_{c}}\lesssim \hbar/(e^{*}V)\ll \hbar/(k_{B}T)$ dominates the integral.
Among the integral domains of $(\tilde{t}_1,\tilde{t}_2)$, we found that the above integral over the domain of $\tilde{t}_{1}+\frac{d}{v_{c}}<0$ and $\tilde{t}_{2}+\frac{d}{v_{c}}<0$ vanishes, while
the integral over the domain of $\tilde{t}_{1}+\frac{d}{v_{c}}>0$ and $\tilde{t}_{2}+\frac{d}{v_{c}}>0$ was already accounted in the computation of the contribution from the time-domain braiding.
We therefore performed the integral over the other domains and found the contribution $C_{2}^{\textrm{interm}}(t_1-t_2)$ for $t_1 > t_2$
\begin{equation} \nonumber
	\begin{split}
		C_{2}^{\textrm{interm}}(t_1-t_2) =	& \,\, i G_{\textrm{M,eq}}^{>}(t_1-t_2)
		\frac{|\xi_\text{iQPC}|^2}{4\pi^{2}a^{2} d_{\sigma}} 
		 \Big[ \frac{ 2\delta_{c}\sin(\pi\delta)\Gamma[1-2\delta]}{\pi (k_{\textrm{B}} T)^2}\Big(\frac{2\pi a k_{\textrm{B}}T}{v_{c}}\Big)^{2\delta_{c}} \Big(\frac{2\pi a k_{\textrm{B}}T}{v_{n}}\Big)^{2\delta_{n}}\Big]\\
		\times & \,\, \textrm{Im} \Big[\frac{\Gamma[\delta-i\frac{e^{*}V_\text{S}}{2\pi k_{\textrm{B}}T}]}{\Gamma[1-\delta-i\frac{e^{*} V_\text{S}}{2\pi k_{\textrm{B}}T}]}\Big(\psi(\delta- i \frac{e^{*} V_\text{S} }{2\pi k_{\textrm{B}}T}) - \psi(1-\delta- i \frac{e^{*} V_\text{S} }{2\pi k_{\textrm{B}}T})\Big)\Big]
	\end{split}
\end{equation} 
where $\psi$ is the diagamma function $\psi(z)=\Gamma[z]'/\Gamma[z]$.
On the other hand, when $t_1 < t_2$, $C_{2}^{\textrm{interm}}(t_1-t_2)$ is nonvanishing under the condition (ii). In a similar manner to the case for $t_1 > t_2$ we obtained $C_{2}^{\textrm{interm}}(t_1-t_2)$ for $t_1 < t_2$, which is identical to the expression for $t_1 > t_2$ up to an overall minus sign.
Combining the results for $t_1 > t_2$ and $t_1 < t_2$, we found the contribution from the lowest-order intermediate process to the $C_2$ term in the charged-anyon configuration on the PH-Pf edge as  
\begin{equation} \label{Eq:Intermediate charged}
	\begin{split} 
		C_{2}^{\textrm{interm}}(t_1-t_2) =	& \,\, i \, \textrm{sgn}(t_1-t_2)  G_{\textrm{M,eq}}^{>}(t_1-t_2)
		R_\text{iQPC} f_\text{interm}^\text{c} (\delta_c,\delta_n,\frac{e^{*}V_\text{S}}{k_\text{B}T}), \\
		f^\text{c}_{\textrm{interm}} (\delta_c,\delta_n,\alpha \equiv \frac{e^{*}V_\text{S}}{k_\text{B}T}) = & \,\, \nu \Big(\frac{e}{e^{*}}\Big)^{2} \frac{\delta_{c}}{\cos(\pi\delta)}
		\frac{\alpha}{2\pi}
		\textrm{csch}\Big(\frac{\alpha}{2}\Big) \textrm{Im} \Big[\sin\Big(\pi\delta + i\frac{\alpha}{2}\Big)\Big(\psi(\delta- i \frac{\alpha}{2\pi} ) - \psi(1-\delta- i \frac{\alpha}{2\pi} )\Big) \Big],
	\end{split}
\end{equation}  
where $\nu=1/2$ and the expression of $R_\text{iQPC}$ in Eq.~\eqref{RiQPC} was used.

We also obtained the contributions from the higher-order intermediate processes to the $C_{2m}$ terms with $m>1$ in Eq.~\eqref{Eq:Keldysh_Formalism PHPf} in the charged-anyon configuration on the PH-Pf edge. In these higher-order processes, a braiding occurs in a manner similar to the time-domain braiding [see Fig.~\ref{fig:intermediate}(b) for the contribution to the $C_4$ term].
Collecting the contributions from all-order intermediate processes, we obtained the Green function
$G_\text{M,neq}^{>}(t_1-t_2) = G_\text{M,neq}^{>, \textrm{TDB}}(t_1-t_2) + G_\text{M,neq}^{>, \textrm{triv}}(t_1-t_2) + G_\text{M,neq}^{>, \textrm{interm}}(t_1-t_2) + \cdots$ for the charged-anyon configuration on the PH-Pf edge:
\begin{equation} \label{eq:interm_correlator}
	\begin{split}
		G_\text{M,neq}^{>, \textrm{interm}}(t_1-t_2)  = & \,\, G_{\textrm{M,eq}}^{>}(t_1-t_2) \, \Delta_{\textrm{interm}}^{\textrm{c}} (t_{1}-t_{2}) \\
        \Delta_{\textrm{interm}}^{\textrm{c}} (t_{1}-t_{2}) = & \,\, i \, \textrm{sgn}(t_1-t_2) R_\text{iQPC} f_\text{interm}^\text{c} (\delta_c,\delta_n,\frac{e^{*}V_\text{S}}{k_\text{B}T}) \Delta_{\textrm{TDB}}^{\textrm{c}} ( t_{1}-t_{2} )
	\end{split}
\end{equation}
where the exponential factor $\Delta_\textrm{TDB}^{\textrm{c}}$ is given by
Eq.~\eqref{eq:TDB_correlator} for the Poissonian injection statistics and by
Eq.~\eqref{eq:exponentialcharge} for the binomial statistics.

For the charged-anyon configuration on the A-Pf edge, we found that the contribution of the intermediate process to the Green function $G_{\text{M,neq;} p_{\text{M}0}}^{>}$ in Eq.~\eqref{Eq:Keldysh_Formalism APf} follows the same expression as Eq.~\eqref{eq:interm_correlator}.  
While the PH-Pf and A-Pf cases have the same expression, their quantitative contributions differ, because the total scaling dimensions $\delta$ of their QPs are $1/4$ for the PH-Pf edge and $1/2$ for the A-Pf edge.

We also studied the case of the neutral-anyon configuration. For the PH-Pf edge, the contribution of the intermediate processes to the Green function $G_\text{M,neq}^{>}$ vanishes; $G_\text{M,neq}^{>,\text{interm}} = \Delta_{\text{interm}}^{\text{n},\textrm{PH-Pf}} =0$. Although the lowest-order intermediate process for a neutral anyon from the diluted beam can arise from four distinct time windows, these contributions cancel for the PH-Pf edge: 
\begin{equation} \nonumber
	\text{(i)} \,\, \tilde{t}_{1}+\frac{d}{v_{n}}\simeq t_2, \,\,\, \tilde{t}_{1}\simeq \tilde{t}_{2},
	\quad \quad  \text{(ii)} \,\, \tilde{t}_{2}+\frac{d}{v_{n}}\simeq t_1,  \,\,\,  \tilde{t}_{1}\simeq \tilde{t}_{2}, \quad \quad  \text{(iii)} \,\, \tilde{t}_{2}+\frac{d}{v_{n}}\simeq t_2,  \,\,\, \tilde{t}_{1}\simeq \tilde{t}_{2},
	\quad \quad  \text{(iv)} \,\, \tilde{t}_{1}+\frac{d}{v_{n}}\simeq t_1,  \,\,\,  \tilde{t}_{1}\simeq \tilde{t}_{2}\,. 
\end{equation} 
This gives twice as many time windows as in the charged-anyon configuration, where the intermediate contribution arises from two time windows. The additional time windows originate from the self-conjugate nature of the Ising anyon $\sigma$, which describes the neutral non-Abelian sector of the PH-Pf QP.
We found that the contribution from the lowest-order intermediate process in the time window (i) exactly cancels with that of (iii), while the contribution of (ii) cancels with that of (iv).

For the A-Pf edge, on the other hand, a diluted neutral-anyon beam injected along the middle edge contains two neutral types, $p_{\rm M i}=\pm1$, as shown in Eqs.~\eqref{Eq:Keldysh_Formalism APf} and \eqref{c2mAPf}. We found that the lowest-order intermediate process associated with a $p_{\rm M i}=p_{\rm M0}$ anyon contributes to $G_{{\rm M,neq};p_{\rm M0}}^{>}(t_1-t_2)$ through the time windows (i) and (ii), whereas the process associated with a $p_{\rm M i}=-p_{\rm M0}$ anyon contributes through the windows (iii) and (iv). These two contributions enter with opposite signs. 
Combining all the contributions from the different time windows, we found that the intermediate contribution to the greater Green function is
\begin{equation}
\label{eq:greenfunintermediateAPfneutral}
    \begin{split}
       G_{\textrm{M,neq},p_{\textrm{M}0}}^{>,\textrm{interm}}(t_{1}-t_{2}) = & \,\, G_\text{M,eq}^{>}(t_1-t_2) \Delta_{\textrm{interm}, p_{\textrm{M}_{0}} }^{\textrm{n},\textrm{A-Pf}} (t_{1}-t_{2}) \\
       \Delta_{\textrm{interm}, p_{\textrm{M}_{0}} }^{\textrm{n},\textrm{A-Pf}} (t_{1}-t_{2}) =& \,\,i\textrm{sgn}(t_{1}-t_{2})  R_\text{iQPC} f^\text{n}_{\textrm{interm}}(\delta_c,\delta_n,\frac{e^{*}V_\text{S}}{k_\text{B}T}) \Delta_{\textrm{TDB}}^{n} (t_1 - t_2) \\
       \times & \Big[\frac{\sum_{p_\text{I} =\pm}|\xi_{\textrm{iQPC},(p_{\textrm{M}_{0}},p_{\text{I}})}|^{2}}{\sum_{\mathbf{p}} |\xi_{\textrm{iQPC},\mathbf{p}}|^{2} } - \frac{\sum_{p_\text{I}=\pm 1}|\xi_{\textrm{iQPC},(-p_{\textrm{M}_{0}},p_{\text{I}})}|^{2}}{\sum_{\mathbf{p}} |\xi_{\textrm{iQPC},\mathbf{p}}|^{2} } \Big] \\
    \end{split}
\end{equation}
Here $f_{\rm interm}^{\rm n}$ is defined analogously to $f_{\rm interm}^{\rm c}$ in Eq.~\eqref{Eq:Intermediate charged}, with $\delta_{c/n}$ in that expression replaced by $\delta_{n/c}$. The exponential factor $\Delta_{\textrm{TDB}}^{\textrm{n}}$ is given by Eq.~\eqref{eq:TDB_correlator neutral2} in the Poissonian limit and by Eq.~\eqref{eq:exponentialneutral} for the binomial statistics.

\subsubsection{Contribution from other subleading braiding processes} \label{subleadbraid}

In the neutral-anyon configuration, subdominant time-domain braiding processes involving the fusion channel $\psi$ and the unit braiding monodromy $M_{I \to \psi}$ can also occur. This should be contrasted with the dominant time-domain braiding involving the vacuum channel $I$ and the monodromy $M_n = M_{I \to I}$ (see \ref{sec_TDB}).
This process was introduced in Fig. S2 of the Supplementary Materials of Ref.~\cite{Lee22}.

For the PH-Pf edge, the contribution of this process to the Green function ($G_\text{M,neq}^{>}$) yields an expression similar to Eq.~(S15) in the Supplementary Materials of Ref.~\cite{Lee22}. This expression vanishes when substituting the scaling dimensions of the PH-Pf edge (the total scaling dimension $\delta = 1/4$ and the Majorana channel $\psi$ dimension $\delta_\psi = 1$) into the expression; see Eq.~(S15) in the Supplementary Materials of Ref.~\cite{Lee22} with the replacement of $h_\psi \to \delta/2$ and $h_a \to \delta_\psi/2$. Namely, this process does not contribute to the Green function. 

For the A-Pf edge, on the other hand, there are two types ($p_{\text{M}i} = \pm 1$) of anyons from a diluted neutral anyon beam along the middle edge, as in Eqs.~\eqref{Eq:Keldysh_Formalism APf} and \eqref{c2mAPf}. 
Consequently, the subdominant time-domain braiding involving the fusion channel $\psi$ has pairs of contributions with opposite signs.
For example,
one process involves an injected anyon with $p_{\text{M}i} = +1$ braiding in time domain with an anyon with $p_{\text{M}0} = +1$ excited at the detection QPC, which is described by the braiding monodromy $M_{I \to \psi} M_{++}$. Here $M_{I \to \psi}$ is the monodromy of the non-Abelian sector, while $M_{++}$ describes the braiding between two Abelian neutral anyons $e^{i p \phi_{n}/2} $ with $p=+1$ [see Eq.~\eqref{APfQP}].
Its partner process involves an injected anyon with $p_{\text{M}i} = -1$ braiding with an anyon with $p_{\text{M}0} = +1$ excited at the detection QPC, described by the monodromy $M_{I \to \psi} M_{-+}$, where $M_{-+}$ accounts for the braiding between two neutral anyons with $p=+1$ and $p=-1$. 
The relation $M_{-+}=-M_{++}$ makes the two partner processes contribute with opposite signs.  
The contribution of these processes to the Green function $G_{\text{M,neq;} p_{\text{M}0}}^{>}$ takes the following form at $t_{1}-t_{2}>0$
\begin{equation}
\label{eq:subleadingGreenfun}
    \begin{split}
        G_{\textrm{M,neq},p_{\textrm{M}0}}^{>,\textrm{subB}} (t_{1}-t_{2}) = & \,\, G_{\textrm{M,eq}}^{>}(t_{1}-t_{2}) f_{\textrm{subB}}(\delta_c,\delta_{n}, \delta_{\psi}, e^{*}V_{\textrm{S}},k_{\textrm{B}} T ,t_{1}-t_{2}) \\
        \times & \frac{M_{I\rightarrow \psi} }{d_{\sigma}} \Big[\frac{\sum_{j=\pm 1}|\xi_{\textrm{iQPC},(p_{\textrm{M}_{0},j})}|^{2}}{\sum_{\mathbf{p}} |\xi_{\textrm{iQPC},\mathbf{p}}|^{2} } \Big(\coth\Big(\frac{e^{*}V_{\textrm{S}}}{2k_{\textrm{B}} T}\Big) \textrm{Re}[M_{ p_{ \textrm{M}0 } p_{ \textrm{M}0 } }] + i \textrm{Im}[M_{p_{\textrm{M}0} p_{ \textrm{M}0 } }]\Big) \\
        +& \frac{\sum_{j=\pm 1}|\xi_{\textrm{iQPC},(-p_{\textrm{M}_{0},j})}|^{2}}{\sum_{\mathbf{p}} |\xi_{\textrm{iQPC},\mathbf{p}}|^{2} } \Big(\coth\Big(\frac{e^{*}V_{\textrm{S}}}{2k_{\textrm{B}} T}\Big) \textrm{Re}[M_{p_{ \textrm{M}0 } (- p_{ \textrm{M}0 }) }] + i \textrm{Im}[M_{p_{ \textrm{M}0 } (- p_{ \textrm{M}0 })} ]\Big)\Big] \\
    \end{split}
\end{equation}
where $f_{\textrm{subB}}(\delta_c,\delta_n,\delta_{\psi},e^*V_{\textrm{S}}, k_\text{B} T,t_{1}-t_{2})$ is given by
\begin{equation}
    \begin{split}
        f_{\textrm{subB}}(\delta_c,\delta_n,\delta_{\psi},e^{*}V_{\textrm{S}},k_{B}T,t_{1}-t_{2}) = &\frac{\sum_{\mathbf{p}} |\xi_{\textrm{iQPC},\mathbf{p}}|^{2}}{2\pi^{2} a v_{c}} \Big(\frac{2\pi a k_{\textrm{B}} T}{v_{c}}\Big)^{2\delta_{c}-1} \Big(\frac{2\pi a k_{\textrm{B}} T}{v_{n}}\Big)^{2\delta_{n}-\frac{\delta_{\psi}}{2}} \Big(\frac{\sin[\pi k_{\textrm{B}} T \{av_{n}^{-1} + i (t_{1}-t_{2})\}]}{\pi k_{\textrm{B}} T a v_{n}^{-1} }\Big)^{\frac{\delta_{\psi}}{2}} \\
        \times & B\Big( \delta-\frac{\delta_{\psi}}{4}+  i\frac{e^{*}V_{\textrm{S}}}{2\pi k_{\textrm{B}}T}, \delta-\frac{\delta_{\psi}}{4} - i\frac{e^{*}V_{\textrm{S}}}{2\pi k_{\textrm{B}}T}\Big) \sinh\Big(\frac{e^{*}V_{\textrm{S}}}{2k_{\textrm{B}}T}\Big)\\
        \times & \int_{0}^{(t_{1}-t_{2})} d\tilde{T} \Big(\frac{\pi k_{\textrm{B}}T a v_{n}^{-1}}{\sin[\pi k_{\textrm{B}}T \{a v_{n}^{-1} + i (\tilde{T}-(t_{1}-t_{2}))\}]}\Big)^{\delta_{\psi}/2} \Big (\frac{ \pi k_{\textrm{B}} T a v_{n}^{-1} }{\sin[\pi k_{\textrm{B}} T \{a v_{n}^{-1} + i \tilde{T} \}]} \Big)^{\delta_{\psi}/2}.
    \end{split}
\end{equation}

\subsection{Tunneling rates and Fano factors} \label{sec_Fano}

For both charged- (c) and neutral-anyon (n) configurations, the tunneling rate from the middle edge (m) to the detection edge (d) and the tunneling rate from the detection edge to the middle edge, respectively, are written in terms of the nonequilibrium Green functions $\mathcal{C}_\text{neq}$ in Eq.~\eqref{neqC} computed in \ref{sec_correlator},
\begin{equation} \label{eq:tunnelingratesQPC}
	\begin{split}
	W_{\text{m} \rightarrow \text{d}}^\text{c/n} & = \int_{-\infty}^{\infty} d(t_1 - t_2) \Big\langle \mathcal{T}_\text{dQPC}^{\dagger}(0) \mathcal{T}_\text{dQPC}(t_1 - t_2) \Big\rangle_{\textrm{neq}}, \\ W_{\text{d} \rightarrow \text{m}}^\text{c/n} & = \int_{-\infty}^{\infty} d(t_1 - t_2) \mathcal{C}_\text{neq} (t_1-t_2)= \int_{-\infty}^{\infty} d(t_1 - t_2) \Big\langle \mathcal{T}_\text{dQPC}(t_1 - t_2) \mathcal{T}_\text{dQPC}^{\dagger}(0) \Big\rangle_{\textrm{neq}}.
	\end{split}
\end{equation}
The tunneling rates are decomposed into those of the time-domain braiding, the trivial partitioning, and the intermediate processes,
\begin{equation} \label{eq:totalrates}
	W_{\text{m} \rightarrow \text{d}}^\text{c/n} \simeq W_{\text{m} \rightarrow \text{d}}^\text{c/n,TDB}+W_{\text{m} \rightarrow \text{d}}^\text{c/n,triv}+W_{\text{m} \rightarrow \text{d}}^\text{c/n,interm}, \quad \quad \quad W_{\text{d} \rightarrow \text{m}}^\text{c/n} \simeq W_{\text{d} \rightarrow \text{m}}^\text{c/n,TDB}+W_{\text{d} \rightarrow \text{m}}^\text{c/n,triv}+W_{\text{d} \rightarrow \text{m}}^\text{c/n,interm}.
\end{equation}
The tunneling current and noise at the detection QPC are written in terms of the tunneling rates as
\begin{equation} \label{eq:currentnoisefromrates}
	I_{ \textrm{dQPC} } = e^{*}(W_{\text{m} \rightarrow \text{d}}^{ \textrm{c/n} } - W_{\text{d} \rightarrow \text{m}}^{\textrm{c/n}} ), \quad \quad \quad S_{ \textrm{dQPC} } = 2(e^{*})^{2}(W_{\text{m} \rightarrow \text{d}}^{\textrm{c/n}} +  W_{\text{d} \rightarrow \text{m}}^{\textrm{c/n} }).
\end{equation}
Substituting these expressions into Eq.~(6) of the main text, we found that the excess noise measured in the detection drain D [D2 in Fig. 1(a) and D1 in Fig. 3(a)] is given by \begin{equation} \label{eq:edgenoisefromrates}
    \begin{split}
        S_{ \textrm{D} } =& S_{\textrm{dQPC}} +2\overline{\Delta I_\text{dQPC}\Delta I_{\text{S}_\text{gnd}}} \\
        =& 2(e^{*})^{2}(W_{\text{m} \rightarrow \text{d}}^{ \textrm{c/n} } +  W_{\text{d} \rightarrow \text{m}}^{\textrm{c/n}} ) - 4 (e^{*})^{2} k_\textrm{B} T\frac{\partial}{\partial (e^{*} V_{\textrm{vir}}) }\big( W_{\text{m} \rightarrow \text{d}}^{\textrm{c/n} , V_{\textrm{vir}} } - W_{\text{d} \rightarrow \text{m}}^{\textrm{c/n} , V_{\textrm{vir} } } \big)\Big|_{V_{\textrm{vir}}=0}.
    \end{split}
\end{equation} 
Here, $V_\textrm{vir}$ is introduced as a small virtual probe voltage to compute the cross correlation in Eq.~(6) of the main text. 
It is incorporated by attaching an additional phase factor to the Green functions of the detection edge, $G_{\text{D}}^{\lessgtr}(t_{1}-t_{2}) \rightarrow e^{ ie^{*}V_{\textrm{vir}}(t_{1}-t_{2})}G_{\text{D}}^{\lessgtr}(t_{1}-t_{2})$. We compute the tunneling rates $W_{\textrm{m/d}\rightarrow \textrm{d/m}}^{V_{\text{vir}}} $ using these modified Green functions, take the derivative with respect to $V_{\text{vir}}$, and then set $V_{\text{vir}} = 0$. The Fano factor for the charged-anyon configuration, defined in Eq.~(3) of the main text, can be expressed in terms of the rates as
\begin{equation} \label{eq:ChargeFano with rate}
	\begin{split}
		\mathcal{F}_\text{charge}= \frac{S_{\text{D}}}{2e^{*}I_{\textrm{dQPC}}} =\frac{(W_{\text{m} \rightarrow \text{d}}^\text{c} + W_{\text{d} \rightarrow \text{m}}^\text{c}) - \left.2 k_\text{B} T \frac{\partial}{\partial e^{*} V_{\textrm{vir}}}(W_{\text{m} \rightarrow \text{d}}^{\text{c},V_{\textrm{vir}}}  - W_{\text{d} \rightarrow \text{m}}^{\text{c},V_{\textrm{vir}}} ) \right|_{V_{\textrm{vir}}=0} }{W_{\text{m} \rightarrow \text{d}}^\text{c} - W_{\text{d} \rightarrow \text{m}}^\text{c}}, 
	\end{split}
\end{equation}

In the neutral-anyon configuration, the differential reflection probability $R_{\textrm{dQPC}}^{\textrm{diff}}$ ($=R_{\textrm{QPC2}}^{\textrm{diff}}$) at the detection QPC was measured experimentally. It can be written as 
\begin{equation}
    R_{\textrm{dQPC}}^{\textrm{diff}} = e^{*}\frac{\partial}{\partial I_{\textrm{ref}}} \big(W_{\text{m} \rightarrow \text{d}}^{\text{n},V_{\textrm{ref}} }- W_{\text{d} \rightarrow \text{m}}^{\text{n}, V_{\textrm{ref}}} \big) \Big|_{I_{\textrm{ref}}\rightarrow 0}.
\end{equation} 
Here, $R_{\textrm{dQPC}}^{\textrm{diff}}$ is defined as the response to a small reference current $ I_{\textrm{ref}}= - (e^{2}/2h)V_{\textrm{ref}}$ induced by the reference AC voltage $V_{\textrm{ref}}$ applied to source S2, while the DC voltage $V_\text{S}$ is applied to source S1.
 The calculation of $R_{\textrm{dQPC}}^{\textrm{diff}}$ is analogous to that of the cross correlation. We emphasize, however that $V_{\text{ref}}$ is an experimentally applied reference voltage, while $V_{\textrm{vir}}$ is a virtual probe voltage introduced only for theoretical computation.
The Fano factor for the neutral-anyon configuration [Eq.~(4) in the main text] is then written as 
\begin{equation} \label{eq:NeutralFano with rate}
	\begin{split}
		\mathcal{F}_\text{neutral}= \frac{S_{\text{D}}}{2e^{*}I_{\textrm{iQPC}}R_{\textrm{dQPC}}^{\textrm{diff}} } =\frac{(W_{\text{m} \rightarrow \text{d}}^\text{n} + W_{\text{d} \rightarrow \text{m}}^\text{n}) - \left. 2 k_\text{B}T \frac{\partial}{\partial e^{*} V_{\textrm{vir}}}(W_{\text{m} \rightarrow \text{d}}^{\text{n},V_{\text{vir}}} - W_{\text{d} \rightarrow \text{m}}^{\text{n},V_{\text{vir}}} )\right|_{V_{\textrm{vir}}=0} }{I_\text{iQPC} \frac{\partial}{\partial I_{\textrm{ref}}}(W_{\text{m} \rightarrow \text{d}}^{\text{n},V_{\textrm{ref}}}  - W_{\text{d} \rightarrow \text{m}}^{\text{n},V_{\textrm{ref}}} )|_{I_{\textrm{ref}}=0} }\,.
	\end{split}
\end{equation}
Below, we present the contribution of each process to the rates.

\subsubsection{Time domain braiding}
Using the results in \ref{sec_TDB}, we obtained the tunneling rates from the time-domain braiding on the PH-Pf edge, 
\begin{equation}
\label{eq:timedomainbraidingrates}
	\begin{split}
		W_{\text{m} \rightarrow \text{d}}^\text{c/n,TDB} = & \frac{W_0^\text{PH-Pf}}{2\pi^{2} \sin(2\pi \delta)}  \textrm{Re}\Big[ e^{i\pi\delta}  B\Big(\delta+\frac{\mathcal{I}_\text{c/n}}{2\pi k_\text{B} T},  \delta -\frac{\mathcal{I}_\text{c/n}}{2\pi k_\text{B} T} \Big) \sin(\pi\delta - \frac{\mathcal{I}_\text{c/n}}{2 k_\text{B} T}) \Big] \\
		W_{\text{d} \rightarrow \text{m}}^\text{c/n,TDB} = & \frac{W_0^\text{PH-Pf}}{2\pi^{2} \sin(2\pi \delta)}  \textrm{Re}\Big[ e^{i\pi\delta} B\Big(\delta+\frac{\mathcal{I}_\text{c/n}^{*}}{2\pi k_\text{B} T},  \delta -\frac{\mathcal{I}_\text{c/n}^{*}}{2\pi k_\text{B} T} \Big) \sin(\pi\delta - \frac{\mathcal{I}_\text{c/n}^{*}}{2 k_\text{B} T}) \Big],
	\end{split}
\end{equation}
where the common factor $W_0^\text{PH-Pf} = \frac{|\xi_{\textrm{dQPC}}|^{2}}{a v_{c} d_{\sigma}} \Big(\frac{2\pi a k_\text{B} T}{v_{c}}\Big)^{2\delta_{c}-1} \Big(\frac{2\pi a k_\text{B} T}{v_{n}} \Big)^{2\delta_{n}}$ is proportional to the equilibrium contribution. The factor
$\mathcal I_\text{c/n}$ denotes the exponents for the charged/neutral-anyon configuration. For the Poissonian injection statistics, $\mathcal I_\text{c}$
and $\mathcal I_\text{n}$ are given by 
\begin{equation} 
\label{eq:exponentspoissonian}
\mathcal{I}_\text{c/n} =\frac{I_\text{iQPC}}{e^{*}}\coth(\frac{e^* V_\text{S}}{2 k_\text{B} T})\text{Re}[1-M_{c/n}]-i\frac{I_\text{iQPC}}{e^{*}}\text{Im}[M_{c/n}].
\end{equation}
while for the binomial statistics they are given by
\begin{align} \label{eq:noneqfactorbinomial}
	\mathcal{I}_{c/n} =  -\frac{I_\text{S} }{e^*} &\Big( \frac{e^{e^* V_\text{S} /k_\text{B} T}}{e^{e^* V_\text{S} /k_\text{B} T} -1} \log \big [1-R_{\text{iQPC}} + R_{\text{iQPC}} M_{c/n}\big ] +\frac{1}{e^{e^* V_\text{S} /k_\text{B} T} -1} 
	\log \big [1-R_{\text{iQPC}} + R_{\text{iQPC}} M_{c/n}^* \big ] \Big) \,.  
\end{align}
The derivative of tunneling rates with respect to $V_{\textrm{vir}}$
in Eqs.~\eqref{eq:ChargeFano with rate} and \eqref{eq:NeutralFano with rate} was found as 
\begin{equation} \label{eq:crosscorrelationTDB}
    \begin{split}
        \frac{\partial}{\partial (e^{*} V_{\textrm{vir}}) }\big(W_{\text{m} \rightarrow \text{d}}^{\text{c/n}, V_{\textrm{vir}} } - W_{\text{d} \rightarrow \text{m}}^{\text{c/n}, V_{\textrm{vir}} } \big)|_{V_{\textrm{vir}}=0} =& \frac{W_{0}^{\textrm{PH-Pf}}}{4\pi^{3} k_{\textrm{B}}T \cos(\pi\delta)}  \textrm{Re} \Big[ B\Big(\delta+\frac{\mathcal{I}_\text{c/n}}{2\pi  k_\text{B} T},  \delta -\frac{\mathcal{I}_\text{c/n}}{2\pi  k_\text{B} T} \Big) \sin(\pi\delta - \frac{\mathcal{I}_\text{c/n}}{2  k_\text{B} T})\\
        \times& \Big(\psi(1-\delta +\frac{\mathcal{I}_\text{c/n}}{2\pi  k_\text{B} T})  - \psi (\delta + \frac{\mathcal{I}_\text{c/n}}{2\pi  k_\text{B} T})\Big)\Big]\;\propto\; \overline{\Delta I_\text{dQPC}\Delta I_{\text{S}_\text{gnd}}}.
    \end{split}
\end{equation}
We note that for the neutral-anyon configuration, the monodromy of the neutral sector vanishes, $M_n = 0$, and thus $\mathcal{I}_{n}$ is purely real, $\textrm{Im} [\mathcal{I}_{n}] =0$. This leads to the fact that the tunneling rates in Eq.~\eqref{eq:timedomainbraidingrates} are equal, $W_{\textrm{m}\rightarrow \textrm{d}}^{\textrm{n},\textrm{TDB}} = W_{\textrm{d}\rightarrow \textrm{m}}^{\textrm{n},\textrm{TDB}}$, and hence the contribution from the time domain braiding to the tunneling current in Eq.~\eqref{eq:currentnoisefromrates} vanishes. This is consistent with the absence of a tunneling current measured in the neutral-anyon configuration.
The derivative of the tunneling rates with respect to $I_{\textrm{ref}}$
in the denominator of Eq.~\eqref{eq:NeutralFano with rate} is given as
\begin{equation} \label{eq:differentialreflectionTDBPHPf}
    \begin{split}
       \left. \frac{\partial}{\partial I_{\textrm{ref}}}(W_{\text{m} \rightarrow \text{d}}^{\text{n},V_{\textrm{ref}}}  - W_{\text{d} \rightarrow \text{m}}^{\text{n},V_{\textrm{ref}}} ) \right|_{I_{\text{ref}}=0} = &  \frac{e^{*}}{ \nu e^{2} } \frac{ W_{0}^{\textrm{PH-Pf}} }{2\pi^{2} k_{B} T \cos(\pi\delta)} \textrm{Re}\Big[B\Big(\delta+\frac{\mathcal{I}_\text{n}}{2\pi  k_\text{B} T},  \delta -\frac{\mathcal{I}_\text{n}}{2\pi  k_\text{B} T} \Big) \sin(\pi\delta - \frac{\mathcal{I}_\text{n}}{2  k_\text{B} T}) \\
        \times & \Big(\psi(1-\delta +\frac{\mathcal{I}_\text{n}}{2\pi  k_\text{B} T})  - \psi (\delta + \frac{\mathcal{I}_\text{n}}{2\pi  k_\text{B} T})\Big) \Big]\;\propto\; R_{\textrm{dQPC}}^{\textrm{diff}}.
    \end{split}
\end{equation}
Equations~\eqref{eq:timedomainbraidingrates} and \eqref{eq:crosscorrelationTDB} provide the time-domain braiding contributions to the tunneling rates and to the cross-correlation, respectively. These quantities enter the total rates in Eq.~\eqref{eq:totalrates} and were used to evaluate the charge Fano factor $\mathcal{F}_{\text{charge}}$ in Eq.~\eqref{eq:ChargeFano with rate} for the PH-Pf edge. Similarly, for the neutral-anyon configuration, Eqs.~\eqref{eq:timedomainbraidingrates}, \eqref{eq:crosscorrelationTDB}, and \eqref{eq:differentialreflectionTDBPHPf} provide the corresponding contributions to the tunneling rates, the cross-correlation, and the differential reflection probability, respectively, which were used 
to evaluate the neutral Fano factor $\mathcal{F}_{\text{neutral}}$ in Eq.~\eqref{eq:NeutralFano with rate}. 

The A-Pf case follows the same expressions as the PH-Pf results, Eqs.~\eqref{eq:timedomainbraidingrates}, \eqref{eq:crosscorrelationTDB}, and \eqref{eq:differentialreflectionTDBPHPf}, but with $W_0^\text{PH-Pf}$ replaced  by $W_0^\text{A-Pf} = \frac{\sum_{\vec{p}}|\xi_{\textrm{dQPC}, \vec{p}}|^{2}}{a v_{c} d_{\sigma}} \Big(\frac{2\pi a k_\text{B} T}{v_{c}}\Big)^{2\delta_{c}-1} \Big(\frac{2\pi a k_\text{B} T}{v_{n}} \Big)^{2\delta_{n}}$. Here, the replacement
$|\xi_{\textrm{dQPC}}|^{2} \rightarrow \sum_{\vec{p}}|\xi_{\textrm{dQPC}, \vec{p}}|^{2} $ accounts for the multiple QP species that can tunnel at the detection QPC for the A-Pf edge.

\subsubsection{Trivial Partitioning}

Using the results in \ref{sec_triv}, we computed the tunneling rates arising from the trivial partitioning.
For the charged-anyon configuration on the PH-Pf edge, the rates read 
\begin{equation} \label{eq:tunnelingratestrivial}
	\begin{split}
		W_{\text{m} \rightarrow \text{d}}^\text{c,triv} = &\frac{W_0^\text{PH-Pf}}{4\pi^{2}}   C_{\textrm{rd}}^{\textrm{c,PH-Pf}} \,R_{\textrm{iQPC}}\, f_{\textrm{triv}}^{\textrm{c}}\big(\delta_c,\delta_n,\frac{e^{*}V_\text{S}}{k_\text{B}T}\big)  B\Big( \delta -i\frac{e^{*}V_\text{S}}{2\pi k_\text{B}T}, \delta +i\frac{e^{*}V_\text{S}}{2\pi k_\text{B}T}\Big) e^{\frac{e^{*}V_\text{S}}{2 k_\text{B}T}}, \\
		 W_{\text{d} \rightarrow \text{m}}^\text{c,triv} = &  \frac{W_0^\text{PH-Pf}}{4\pi^{2}}    C_{\textrm{rd}}^{\textrm{c,PH-Pf}} \,R_{\textrm{iQPC}}\, f_{\textrm{triv}}^{\textrm{c}}\big(\delta_c,\delta_n,\frac{e^{*}V_\text{S}}{k_\text{B} T}\big) B\Big( \delta -i\frac{e^{*}V_\text{S}}{2\pi k_\text{B}T}, \delta +i\frac{e^{*}V_\text{S}}{2\pi k_\text{B}T}\Big) e^{-\frac{e^{*}V_\text{S}}{2 k_\text{B}T}}.
	\end{split}
\end{equation}
 The derivative of the tunneling rates in Eq.~\eqref{eq:ChargeFano with rate} is expressed as 
\begin{equation} \label{eq:crosscorrelationChargetriv}
    \begin{split}
        \left. \frac{\partial}{\partial (e^{*} V_{\textrm{vir}}) }\big(W_{\text{m} \rightarrow \text{d}}^{\text{c}, V_{\textrm{vir}} } -  W_{\text{d} \rightarrow \text{m}}^{\text{c}, V_{\textrm{vir}} } \big) \right|_{V_{\textrm{vir}}=0} = \frac{W_{0}^{\textrm{PH-Pf}}}{4 \pi^{2} k_{\textrm{B}} T}  C_{\textrm{rd}}^{\textrm{c,PH-Pf}} \,R_{\textrm{iQPC}}\, f_{\textrm{triv}}^{\textrm{c}}\big(\delta_c,\delta_n,\frac{e^{*}V_\text{S}}{k_\text{B}T}\big) B\Big( \delta -i\frac{e^{*}V_\text{S}}{2\pi k_\text{B}T}, \delta +i\frac{e^{*}V_\text{S}}{2\pi k_\text{B}T}\Big) \\
        \times  \Big( \cosh\Big(\frac{e^{*}V_{S}}{2k_\textrm{B} T}\Big) - \frac{2}{\pi}\sinh\Big(\frac{e^{*}V_{S}}{2k_\textrm{B} T}\Big) \Im\Big[\psi(\delta + i \frac{e^{*}V_{S}}{2 \pi k_{\textrm{B} }T} )\Big] \Big)\;\propto\; \overline{\Delta I_\text{dQPC}\Delta I_{\text{S}_\text{gnd}}}.
    \end{split}
\end{equation}
Equations~\eqref{eq:tunnelingratestrivial} and \eqref{eq:crosscorrelationChargetriv} provide the trivial-partitioning contributions to the tunneling rates and to the cross-correlation, respectively. These quantities enter the total rates in Eq.~\eqref{eq:totalrates} and were used to evaluate the charge Fano factor $\mathcal{F}_{\text{charge}}$ in Eq.~\eqref{eq:ChargeFano with rate} for the PH-Pf edge.

For the charged-anyon configuration on the A-Pf edge, the results follow the expressions in Eqs.~\eqref{eq:tunnelingratestrivial} and \eqref{eq:crosscorrelationChargetriv} for PH-Pf edge, but with $W_0^\text{PH-Pf}$ replaced by $W_0^\text{A-Pf} = \frac{\sum_{\vec{p}}|\xi_{\textrm{dQPC}, \vec{p}}|^{2}}{a v_{c} d_{\sigma}} \Big(\frac{2\pi a k_\text{B} T}{v_{c}}\Big)^{2\delta_{c}-1} \Big(\frac{2\pi a k_\text{B} T}{v_{n}} \Big)^{2\delta_{n}}$.

For the neutral-anyon configuration on the PH-Pf edge, we obtained the tunneling rates from the trivial partitioning,
\begin{equation} \label{Eq:Neutral_Triv_rate}
	\begin{split}
		W_{\text{m} \rightarrow \text{d}}^\text{n,triv}= & \frac{ W_0^\text{PH-Pf}}{4\pi^{2}} 
		C_{\textrm{rd}}^{\textrm{n,PH-Pf}} \;p_{\textrm{m}\rightarrow\textrm{d}}^{\textrm{PH-Pf}} \,R_{\textrm{iQPC}}\, f_{\textrm{triv}}^{\textrm{n}}\big(\delta_c,\delta_n,\frac{e^{*}V_\text{S}}{k_\text{B}T},d_{\sigma} \big)  B\Big( \delta -i\frac{e^{*}V_\text{S}}{2\pi k_\text{B}T}, \delta +i\frac{e^{*}V_\text{S}}{2\pi k_\text{B}T}\Big) \cosh\Big(\frac{e^{*}V_\text{S}}{2k_\text{B}T}\Big), \\
        W_{\text{d} \rightarrow \text{m}}^\text{n,triv}= & \frac{ W_0^\text{PH-Pf}}{4\pi^{2}} 
		C_{\textrm{rd}}^{\textrm{n,PH-Pf}} \;p_{\textrm{d}\rightarrow\textrm{m}}^{\textrm{PH-Pf}} \,R_{\textrm{iQPC}}\, f_{\textrm{triv}}^{\textrm{n}}\big(\delta_c,\delta_n,\frac{e^{*}V_\text{S}}{k_\text{B}T},d_{\sigma} \big)  B\Big( \delta -i\frac{e^{*}V_\text{S}}{2\pi k_\text{B}T}, \delta +i\frac{e^{*}V_\text{S}}{2\pi k_\text{B}T}\Big) \cosh\Big(\frac{e^{*}V_\text{S}}{2k_\text{B}T}\Big), \\
	\end{split}
\end{equation}
with the factor $p_{\textrm{m}\rightarrow\textrm{d}}^{\textrm{PH-Pf}} = p_{\textrm{d}\rightarrow\textrm{m}}^{\textrm{PH-Pf}} =p^{\textrm{PH-Pf}} = 2$. For the PH-Pf edge, $ W_{\text{m} \rightarrow \text{d}}^\text{n,triv} = W_{\text{d} \rightarrow \text{m}}^\text{n,triv} $ and hence the resulting tunneling current $I_{ \textrm{dQPC} }$ at the detection QPC [Eq.~\eqref{eq:currentnoisefromrates}] vanishes. 
This result of $ W_{\text{m} \rightarrow \text{d}}^\text{n,triv} = W_{\text{d} \rightarrow \text{m}}^\text{n,triv} $ originates from the particle-hole-symmetric structure of Ising anyons.
The calculation of the derivative of the tunneling rates in Eq.~\eqref{eq:NeutralFano with rate} leads to
\begin{equation} \label{eq:crosscorrelationtrivPHPf}
    \begin{split}
       \left. \frac{\partial}{\partial (e^{*} V_{\textrm{vir}}) } \big(W_{\text{m} \rightarrow \text{d}}^{\text{n}, V_{\textrm{vir}} } - W_{\text{d} \rightarrow \text{m}}^{\text{n}, V_{\textrm{vir}} } \big)\right|_{V_{\textrm{vir}}=0}&= \frac{W_{0}^{\textrm{PH-Pf}}}{4 \pi^{2} k_{\textrm{B}} T}  C_{\textrm{rd}}^{\textrm{n,PH-Pf}} \,p^{\textrm{PH-Pf}} \,R_{\textrm{iQPC}}\, f_{\textrm{triv}}^{\textrm{n}}\big(\delta_c,\delta_n,\frac{e^{*}V_\text{S}}{k_\text{B}T}, d_{\sigma} \big)  B\Big( \delta -i\frac{e^{*}V_\text{S}}{2\pi k_\text{B}T}, \delta +i\frac{e^{*}V_\text{S}}{2\pi k_\text{B}T}\Big) \\
         & \times  \Big( \cosh\Big(\frac{e^{*}V_{S}}{2k_\textrm{B} T}\Big) - \frac{2}{\pi}\sinh\Big(\frac{e^{*}V_{\textrm{S}}}{2k_\textrm{B} T}\Big) \Im\Big[\psi(\delta + i \frac{e^{*}V_{\textrm{S}}}{2 \pi k_{\textrm{B} }T}) \Big]\Big).
    \end{split}
\end{equation}
The derivative of the rates with respect to $I_{\textrm{ref}}$ in the denominator of Eq.~\eqref{eq:NeutralFano with rate} is given by 
\begin{equation} \label{eq:differentialreflectiontrivPHPf}
    \begin{split}
       \left. \frac{\partial}{\partial I_{\textrm{ref}}}(W_{\text{m} \rightarrow \text{d}}^{\text{n},V_{\textrm{ref}}}  - W_{\text{d} \rightarrow \text{m}}^{\text{n},V_{\textrm{ref}}} )\right|_{I_{\text{ref}}= 0} = &  \frac{e^{*}}{\nu e^{2}} \frac{W_{0}^{\textrm{PH-Pf}}}{2 \pi k_{\textrm{B} }T }C_{\textrm{rd}}^{\textrm{n,PH-Pf}} \;p^{\textrm{PH-Pf}}\,R_{\textrm{iQPC}}\, f_{\textrm{triv}}^{\textrm{n}}\big(\delta_c,\delta_n,\frac{e^{*}V_\text{S}}{k_\text{B}T}, d_{\sigma} \big)\\
		\times & \Big[\cosh\Big(\frac{ e^{*} V_\text{S}}{2k_\text{B}T}\Big) -\frac{2}{\pi}\textrm{Im} \Big[\psi(\delta + i\frac{ e^{*}V_\text{S} }{2\pi k_\text{B}T})\Big] \sinh\Big(\frac{ e^{*} V_\text{S}}{2k_\text{B}T}\Big) \Big]\;\propto\;R_{\textrm{dQPC}}^{\textrm{diff}} .
    \end{split}
\end{equation}
Equations~\eqref{Eq:Neutral_Triv_rate}, \eqref{eq:crosscorrelationtrivPHPf}, and \eqref{eq:differentialreflectiontrivPHPf} provide the trivial-partitioning contributions to the tunneling rates, the cross-correlation, and the differential reflection probability, respectively, which were used 
to evaluate the neutral Fano factor $\mathcal{F}_{\text{neutral}}$ in Eq.~\eqref{eq:NeutralFano with rate}.

For the neutral-anyon configuration on the A-Pf edge, we obtained the same results for the tunneling rates as those for the PH-Pf case in Eq.~\eqref{Eq:Neutral_Triv_rate}, 
but with the replacement of $W_0^\text{PH-Pf} \rightarrow W_0^\text{A-Pf}$, $C_{\textrm{rd}}^{\textrm{n,PH-Pf}} \rightarrow C_{\textrm{rd}}^{\textrm{n,A-Pf}}$, $p_{\text{m}\rightarrow \text{d}}^{\text{PH-Pf}} \rightarrow p_{\text{m}\rightarrow \text{d}}^{\text{A-Pf}}$, and $p_{\text{d}\rightarrow \text{m}}^{\text{PH-Pf}} \rightarrow p_{\text{d}\rightarrow \text{m}}^{\text{A-Pf}}$. 
Specifically, $p_{\textrm{m}\rightarrow\textrm{d}}^{\textrm{A-Pf}}$ and $p_{\textrm{d}\rightarrow\textrm{m}}^{\textrm{A-Pf}}$ are given by
\begin{subequations}
    \begin{align} \label{Eq:A-Pf_triv_mtod}
        p_{\text{m}\rightarrow \text{d}}^{\textrm{A-Pf}} = & \frac{1}{\cosh(\frac{e^{*}V_{\textrm{S}}}{2k_{\textrm{B}}T})}\Bigg[\frac{\sum\limits_{p_{\textrm{D}}=\pm} |\xi_{\textrm{dQPC},(p_{\textrm{D}},p_{\textrm{M}}=+1)}|^{2} }{ \sum\limits_{\mathbf{p}} |\xi_{\textrm{dQPC},\mathbf{p}}|^{2} } \Big(\frac{\sum\limits_{p_{\textrm{I}}=\pm 1}|\xi_{\textrm{iQPC},(p_{\textrm{M}}=+1,p_{\textrm{I}})}|^{2}}{\sum\limits_{\mathbf{p}} |\xi_{\textrm{iQPC},\mathbf{p}}|^{2} } e^{\frac{e^{*}V_{\textrm{S}}}{2k_{\textrm{B}} T}} + \frac{\sum\limits_{p_{\textrm{I}}=\pm}|\xi_{\textrm{iQPC},(p_{\textrm{M}}=-1,p_{\textrm{I}})}|^{2}}{\sum\limits_{\mathbf{p}} |\xi_{\textrm{iQPC},\mathbf{p}}|^{2} }e^{-\frac{e^{*}V_{\textrm{S}}}{2k_{\textrm{B}}T}} \Big) \nonumber \\
        +& \frac{\sum\limits_{p_{\textrm{D}} = \pm}   |\xi_{\textrm{dQPC},(p_{\textrm{D}},p_{\textrm{M}}=-1)}|^{2} }{ \sum\limits_{\mathbf{p}} |\xi_{\textrm{dQPC},\mathbf{p}}|^{2} } \Big(\frac{\sum\limits_{p_{\textrm{I}}=\pm 1}|\xi_{\textrm{iQPC},(p_{\textrm{M}}=+1,p_{\textrm{I}})}|^{2}}{\sum\limits_{\mathbf{p}} |\xi_{\textrm{iQPC},\mathbf{p}}|^{2} } e^{-\frac{e^{*}V_{\textrm{S}}}{2k_{\textrm{B}}T}} + \frac{\sum\limits_{p_{\textrm{I}}=\pm}|\xi_{\textrm{iQPC},(p_{\textrm{M}}=-1,p_{\textrm{I}})}|^{2}}{\sum\limits_{\mathbf{p}} |\xi_{\textrm{iQPC},\mathbf{p}}|^{2} }e^{\frac{e^{*}V_{\textrm{S}}}{2k_{\textrm{B}}T}} \Big)\Bigg] \\
        p_{\text{d}\rightarrow \text{m}}^{\textrm{A-Pf}} = & \frac{1}{\cosh(\frac{e^{*}V_{S}}{2k_{\textrm{B}}T})} \Bigg[\frac{\sum\limits_{p_{\textrm{D}} = \pm} |\xi_{\textrm{dQPC},(p_{\textrm{D}},p_{\textrm{M}}=+1)}|^{2} }{ \sum\limits_{\mathbf{p}} |\xi_{\textrm{dQPC},\mathbf{p}}|^{2} } \Big(\frac{\sum\limits_{p_{\textrm{I}}=\pm 1}|\xi_{\textrm{iQPC},(p_{\textrm{M}}=+1,p_{\textrm{I}})}|^{2}}{\sum\limits_{\mathbf{p}} |\xi_{\textrm{iQPC},\mathbf{p}}|^{2} } e^{-\frac{e^{*}V_{\textrm{S}}}{2k_{\textrm{B}}T}} + \frac{\sum\limits_{p_{\textrm{I}}=\pm}|\xi_{\textrm{iQPC},(p_{\textrm{M}}=-1,p_{\textrm{I}})}|^{2}}{\sum\limits_{\mathbf{p}} |\xi_{\textrm{iQPC},\mathbf{p}}|^{2} }e^{\frac{e^{*}V_{\textrm{S}}}{2k_{\textrm{B}}T}} \Big)\nonumber  \\
        +& \frac{\sum\limits_{p_{\textrm{D}} = \pm} |\xi_{\textrm{dQPC},(p_{\textrm{D}},p_{\textrm{M}}=-1)}|^{2} }{ \sum\limits_{\mathbf{p}} |\xi_{\textrm{dQPC},\mathbf{p}}|^{2} } \Big(\frac{\sum\limits_{p_{\textrm{I}}=\pm 1}|\xi_{\textrm{iQPC},(p_{\textrm{M}}=+1,p_{\textrm{I}})}|^{2}}{\sum\limits_{\mathbf{p}} |\xi_{\textrm{iQPC},\mathbf{p}}|^{2} } e^{\frac{e^{*}V_{\textrm{S}}}{2k_{\textrm{B}}T}} + \frac{\sum\limits_{p_{\textrm{I}}=\pm}|\xi_{\textrm{iQPC},(p_{\textrm{M}}=-1,p_{\textrm{I}})}|^{2}}{\sum\limits_{\mathbf{p}} |\xi_{\textrm{iQPC},\mathbf{p}}|^{2} }e^{-\frac{e^{*}V_{\textrm{S}}}{2k_{\textrm{B}}T}} \Big)\Bigg]\,.
    \label{Eq:A-Pf_triv_dtom}
    \end{align}
\end{subequations}
In contrast to the PH–Pf edge, the vanishing tunneling current arising from the trivial-partitioning contribution is not guaranteed in general for the A–Pf edge.
As can be seen in Eq.~\eqref{Eq:Neutral_Triv_rate}, the tunneling current vanishes only when $p_{\text{m}\rightarrow \text{d}}^{\textrm{A-Pf}} = p_{\text{d}\rightarrow \text{m}}^{\textrm{A-Pf}} $. This is satisfied if one of the following two conditions is satisfied:
\begin{subequations}
\label{eq:vanishingcurrentcond}
\begin{align} \label{eq:vanishingcurrentcond1}
   \text{condition (i)} \quad &\sum_{p_{\textrm{I}} = \pm} |\xi_{\text{iQPC}, ( p_{\text{M}}=1,p_{\textrm{I}})}|^2 = \sum_{p_{\textrm{I}} = \pm}|\xi_{\text{iQPC}, (p_{\text{M}}=-1, p_{\textrm{I}} )}|^2  \,, \\ 
   \text{condition (ii)} \quad & \sum_{p_{\textrm{D}} = \pm} |\xi_{\text{dQPC}, (p_{\textrm{D}}, p_{\text{M}}=1)}|^2 = \sum_{p_{\textrm{D}} = \pm}|\xi_{\text{dQPC}, (p_{\textrm{D}}, p_{\text{M}}=-1 )}|^2\,.
\label{eq:vanishingcurrentcond2}
\end{align}
\end{subequations}
Physically, for a given injected neutral anyon type, the same-neutral-type tunneling at the detection QPC provides the dominant contribution to the ${\rm m}\to{\rm d}$ current ($\propto p_{\text{m} \to \text{d}}^{\text{A-Pf}}$) in the limit of $e^* V_{\textrm{S}} \gg k_\text{B} T$, whereas the opposite-type process provides the dominant contribution to the reverse ${\rm d}\to{\rm m}$ current ($\propto p_{\text{d} \to \text{m}}^{\text{A-Pf}}$) with the same magnitude.
Therefore, the net tunneling current vanishes if the two types are weighted equally, either in the injected beam [condition (i)] or in the tunneling amplitudes at the detection QPC [condition (ii)].
In our experiments, no tunneling current was observed at the detection QPC in the neutral-anyon configuration. This implies that at least one of these conditions in Eq.~\eqref{eq:vanishingcurrentcond} was satisfied in our experiments. In either case, $p_{\textrm{m}\rightarrow \textrm{d}}^{\textrm{A-Pf}}=p_{\textrm{d}\rightarrow \textrm{m}}^{\textrm{A-Pf}}=1$. This leads to $W_{\text{m} \rightarrow \text{d}}^\text{n,triv} = W_{\text{d} \rightarrow \text{m}}^\text{n,triv}$, and hence the trivial partitioning does not contribute to the tunneling current $I_{ \textrm{dQPC} }$ at the detection QPC [Eq.~\eqref{eq:currentnoisefromrates}]. 
Under this zero-current condition for the A-Pf edge, we use Eqs.~\eqref{eq:crosscorrelationtrivPHPf}, and \eqref{eq:differentialreflectiontrivPHPf}, with substituting $W_0^\text{PH-Pf} \rightarrow W_0^\text{A-Pf}$, $C_{\textrm{rd}}^{\textrm{n,PH-Pf}} \rightarrow C_{\textrm{rd}}^{\textrm{n,A-Pf}}$, $p_{\text{m}\rightarrow \text{d}}^{\text{PH-Pf}} \rightarrow p_{\text{m}\rightarrow \text{d}}^{\text{A-Pf}}$, and $p^{\text{PH-Pf}}=2 \rightarrow p^{\text{A-Pf}}=1$.

Finally we discuss about the following point. In the regime of $R_\text{iQPC} \ll 1$ at zero temperature,
the contribution of the trivial partitioning to the tunneling current and noise across the detection QPC is sub-dominant compared to that of the braiding process as
$\frac{W^\text{triv}}{W^\text{TDB}} \propto R_\text{iQPC}^{2-2\delta}$, supporting Eq.~\eqref{triv vs TDB}.

\subsubsection{Intermediate Process}

Using the results in \ref{sec_interm}, we computed the tunneling rates from the intermediate processes for the charged-anyon configuration on PH-Pf edge, 
\begin{equation}
\label{eq:tunnelingrateintermed}
	\begin{split}
		W_{\text{m} \rightarrow \text{d}}^{\textrm{c}, \textrm{interm}}= & \frac{  W_0^\text{PH-Pf}}{4\pi^{2} }  R_{\textrm{iQPC}}\, f_{\textrm{interm}}^{\textrm{c}}\big(\delta_c,\delta_n, \frac{e^{*}V_\text{S}}{k_\text{B}T}\big) \\
		\times& \frac{\pi}{\Gamma[2\delta]}\Big( -\frac{1}{\cos(\pi\delta)} \textrm{Re}\Big[\frac{ \Gamma[\delta + \frac{\mathcal{I}_\text{c}}{2\pi k_\text{B} T}] }{ \Gamma[1-\delta + \frac{\mathcal{I}_\text{c}}{2\pi k_\text{B} T}] }\Big] - \frac{1}{\sin(\pi\delta)} \textrm{Im}\Big[\frac{ \Gamma[\delta + \frac{\mathcal{I}_\text{c}}{2\pi k_\text{B} T}] }{ \Gamma[1-\delta + \frac{\mathcal{I}_\text{c}}{2\pi k_\text{B} T}] }\Big]  +\frac{1}{1-2\delta} \Big(\frac{2
        \pi T a}{\overline{v}}\Big)^{1-2\delta } \Big), \\
		W_{\text{d} \rightarrow \text{m}}^{\textrm{c}, \textrm{interm}}= & \frac{ W_0^\text{PH-Pf}}{4\pi^{2} }  R_{\textrm{iQPC}}\, f_{\textrm{interm}}^{\textrm{c}}\big(\delta_c,\delta_n,\frac{e^{*}V_\text{S}}{k_\text{B}T}\big) \\
		\times& \frac{\pi}{\Gamma[2\delta]}\Big( \frac{1}{\cos(\pi\delta)} \textrm{Re}\Big[\frac{ \Gamma[\delta + \frac{\mathcal{I}_\text{c}}{2\pi k_\text{B} T}] }{ \Gamma[1-\delta + \frac{\mathcal{I}_\text{c}}{2\pi k_\text{B} T}] }\Big]- \frac{1}{\sin(\pi\delta)} \textrm{Im}\Big[\frac{ \Gamma[\delta + \frac{\mathcal{I}_\text{c}}{2\pi k_\text{B} T}] }{ \Gamma[1-\delta + \frac{\mathcal{I}_\text{c}}{2\pi  k_\text{B} T}] }\Big]  -\frac{1}{1-2\delta} \Big(\frac{2
        \pi T a}{\overline{v}}\Big)^{1-2\delta }\Big).
	\end{split}
\end{equation} where the exponential factor $\mathcal I_c$ follows Eqs.~\eqref{eq:exponentspoissonian} for the Poissonian injection statistics and Eqs.~\eqref{eq:noneqfactorbinomial} for the binomial statistics. 
We note that the large parenthesis in the expression for the tunneling rates includes three terms. The first and second terms originate from the long-time part of the time integral, with the characteristic time scale set by $t\sim 1/(R_{\text{iQPC}} |e^* V_{\textrm{S}}|)$, while the last terms arises from short times $\sim a/\overline{v}$ where $\overline{v}$ is a characteristic velocity associated with $v_c$ and $v_n$. For the PH-Pf edge with $\delta = 1/4 < 1/2$, the last term vanishes in the limit of $a\rightarrow 0$ and does not contributes to the tunneling rates.  The derivative of tunneling rates with respect to $V_{\textrm{vir}}$ in Eq.~\eqref{eq:ChargeFano with rate}
\begin{equation} \label{eq:crosscorrelationInterm}
    \begin{split}
        \frac{\partial}{\partial (e^{*} V_{\textrm{vir}}) }\big(W_{\text{m} \rightarrow \text{d}}^{\text{c}, V_{\textrm{vir}} } - & W_{\text{d} \rightarrow \text{m}}^{\text{c}, V_{\textrm{vir}} } \big)|_{V_{\textrm{vir}}=0} = \frac{W_{0}^{\textrm{PH-Pf}}}{4\pi^{3} k_{\textrm{B}}T \cos(\pi\delta)} R_{\textrm{iQPC}} f_{\textrm{interm}}^{\textrm{c}}\big(\delta_c,\delta_n,\frac{e^{*}V_\text{S}}{k_\text{B}T}\big) \\
        \times &\textrm{Im} \Big[B\Big( \delta +\frac{\mathcal{I}_\text{c}}{2\pi k_\text{B} T} , \delta - \frac{\mathcal{I}_\text{c}}{2\pi k_\text{B} T}  \Big) \sin\big(\pi\delta-\frac{\mathcal{I}_\text{c}}{2 k_\text{B} T}\big) \Big( \psi(\delta+\frac{\mathcal{I}_\text{c}}{2\pi k_\text{B} T} \big)- \psi(1-\delta+\frac{\mathcal{I}_\text{c}}{2\pi k_\text{B} T} \big) \Big)\Big] .
    \end{split}
\end{equation}
Equations~\eqref{eq:tunnelingrateintermed} (with the vanishing second term) and \eqref{eq:crosscorrelationInterm} provide the contribution of the intermediate process to the tunneling rates and to the cross-correlation, respectively. These quantities enter the total rates in Eq.~\eqref{eq:totalrates} and were used to evaluate the charge Fano factor $\mathcal{F}_{\text{charge}}$ in Eq.~\eqref{eq:ChargeFano with rate} for the PH-Pf edge.

For the charged-anyon configuration on the A-Pf edge, the rates follow Eqs.~\eqref{eq:tunnelingrateintermed} and \eqref{eq:crosscorrelationInterm} with $W_0^{\rm PH\text{-}Pf}$ replaced by $W_0^{\rm A\text{-}Pf}$. 
For the A-Pf edge, the short-time contribution is important, since $\delta=1/2$: The last term in the large parentheses of Eq.~\eqref{eq:tunnelingrateintermed} has the same power of the cutoff $a$ as the first and second terms and therefore cannot be neglected. In particular, this short-time term is required to cancel the apparent divergence of the long-time contribution. More explicitly, in the the expression of the tunneling current ($\propto W_{\rm m\to d}^{\rm interm}-W_{\rm d\to m}^{\rm interm}$), the divergent part of the first term is exactly cancelled by the short-time cutoff contribution. This guarantees that the intermediate-process contribution to the tunneling current at the detection QPC vanishes for the A-Pf edge, while it has a finite contribution to the noise.

For the neutral-anyon configuration on the PH-Pf edge, the intermediate-process contribution vanishes identically at the level of the Green function $G_{\rm M,neq}^{>,{\rm interm}}$, and hence gives no contribution to the tunneling rates. 
\begin{equation}
\label{eq:tunnelingrateintermedneutral}
	\begin{split}
		W_{\text{m} \rightarrow \text{d}}^{ \textrm{n}, \textrm{interm}}= & 0,\quad \textrm{for the PH-Pf} \\
		W_{\text{d} \rightarrow \text{m}}^{ \textrm{n}, \textrm{interm}}= & 0,\quad \textrm{for the PH-Pf} 
	\end{split}
\end{equation} 
Therefore the intermediate process does not enter transport observables and hence the neutral Fano factor $\mathcal{F}_{\text{neutral}}$ in the neutral-anyon configuration on the PH-Pf edge.

For the neutral-anyon configuration on the A-Pf edge, the corresponding Green function, Eq.~\eqref{eq:greenfunintermediateAPfneutral}, is not identically zero in general, leading to finite tunneling rates, which have the same form as Eqs.~\eqref{eq:tunnelingrateintermed}, but with the replacements of $W_{0}^{\text{PH-Pf}} \rightarrow W_{0}^{\text{A-Pf}}$, $\delta_{c/n}\rightarrow \delta_{n/c}$, and $\mathcal{I}_{c}\rightarrow \mathcal{I}_{n} $. Furthermore, the rates are proportional to 
\begin{subequations} 
\label{eq:tunnelingrateintermed_n}
    \begin{align} \label{eq:tunnelingrateintermed_n_1}
 W_{\text{m}\rightarrow \text{d}}^{\textrm{n,interm} } \propto & \Big(\frac{\sum_{p_{\textrm{I}}=\pm 1}|\xi_{\textrm{iQPC},(p_{\textrm{M}}=+1,p_{\textrm{I}})}|^{2}}{\sum_{\mathbf{p}} |\xi_{\textrm{iQPC},\mathbf{p}}|^{2} }- \frac{\sum_{p_{\textrm{I}}=\pm}|\xi_{\textrm{iQPC},(p_{\textrm{M}}=-1,p_{\textrm{I}})}|^{2}}{\sum_{\mathbf{p}} |\xi_{\textrm{iQPC},\mathbf{p}}|^{2} } \Big) \nonumber\\
        \times & \Big(\frac{\sum_{p_{\textrm{D}}=\pm} |\xi_{\textrm{dQPC},(p_{\textrm{D}},p_{\textrm{M}}=+1)}|^{2} }{ \sum_{\mathbf{p}} |\xi_{\textrm{dQPC},\mathbf{p}}|^{2} } -\frac{\sum_{p_{\textrm{D}} = \pm}   |\xi_{\textrm{dQPC},(p_{\textrm{D}},p_{\textrm{M}}=-1)}|^{2} }{ \sum_{\mathbf{p}} |\xi_{\textrm{dQPC},\mathbf{p}}|^{2} }\Big),  \quad \textrm{for the A-Pf} \\
 W_{\text{d}\rightarrow \text{m}}^{\textrm{n,interm}} \propto & \Big(\frac{\sum_{p_{\textrm{I}}=\pm 1}|\xi_{\textrm{iQPC},(p_{\textrm{M}}=+1,p_{\textrm{I}})}|^{2}}{\sum_{\mathbf{p}} |\xi_{\textrm{iQPC},\mathbf{p}}|^{2} }- \frac{\sum_{p_{\textrm{I}}=\pm}|\xi_{\textrm{iQPC},(p_{\textrm{M}}=-1,p_{\textrm{I}})}|^{2}}{\sum_{\mathbf{p}} |\xi_{\textrm{iQPC},\mathbf{p}}|^{2} } \Big) \nonumber \\
        \times & \Big(\frac{\sum_{p_{\textrm{D}}=\pm} |\xi_{\textrm{dQPC},(p_{\textrm{D}},p_{\textrm{M}}=+1)}|^{2} }{ \sum_{\mathbf{p}} |\xi_{\textrm{dQPC},\mathbf{p}}|^{2} } -\frac{\sum_{p_{\textrm{D}} = \pm}   |\xi_{\textrm{dQPC},(p_{\textrm{D}},p_{\textrm{M}}=-1)}|^{2} }{ \sum_{\mathbf{p}} |\xi_{\textrm{dQPC},\mathbf{p}}|^{2} }\Big), \quad \textrm{for the A-Pf}\,. 
    \label{eq:tunnelingrateintermed_n_2}
    \end{align}
\end{subequations}
However, the resulting intermediate-process contribution to the tunneling rates vanishes, $W_{\text{m}\rightarrow \text{d}}^{\textrm{interm} } = W_{\text{d}\rightarrow \text{m}}^{\textrm{interm} }= 0$, whenever either of the symmetry conditions in Eqs.~\eqref{eq:vanishingcurrentcond} is satisfied. This follows directly from the product of the two parenthesis in the rates~\eqref{eq:tunnelingrateintermed_n}: Eq.~\eqref{eq:vanishingcurrentcond1} sets the first parenthesis to zero, whereas Eq.~\eqref{eq:vanishingcurrentcond2} sets the second parenthesis to zero.
Hence, under either of these symmetry conditions, the intermediate process does not contribute to the transport observables, including the neutral Fano factor $\mathcal{F}_{\text{neutral}}$, for the A-Pf edge. 
As discussed earlier, our experimental observation of a vanishing tunneling current at the detection QPC in the neutral-anyon configuration implies that one of these symmetry conditions is satisfied. Accordingly, for the A-Pf edge, the intermediate process does not contribute to transport in the neutral-anyon configuration realized in our experiments.

In the regime of $R_\text{iQPC} \ll 1$ at zero temperature,
the contribution of the trivial partitioning process to the tunneling current and noise across the detection QPC becomes sub-dominant compared to that of the braiding process as
$\frac{W^\text{interm}}{W^\text{TDB}} \propto R_\text{iQPC}$. It supports Eq.~\eqref{interm vs TDB}.

\subsubsection{Subleading braiding process}

We finally comment on the subleading braiding process discussed in \ref{subleadbraid}. For the neutral-anyon configuration on the PH-Pf edge, the subleading-braiding contribution to the nonequilibrium Green function vanishes identically, as discussed in \ref{subleadbraid}. Therefore, this process does not contribute to the tunneling rates, the tunneling current, or the noise. 
\begin{equation}
\label{eq:tunnelingratesub}
	\begin{split}
		W_{\text{m} \rightarrow \text{d}}^{ \textrm{n}, \textrm{sub}}= & 0,\quad \textrm{for the PH-Pf} \\
		W_{\text{d} \rightarrow \text{m}}^{ \textrm{n}, \textrm{sub}}= & 0,\quad \textrm{for the PH-Pf} 
	\end{split}
\end{equation}

For the A-Pf edge, the corresponding contribution to the Green function is in general finite, when the two neutral anyon types enter with different tunneling coefficients $|\xi_{\text{QPC},\vec{p}}|^2$ [see Eq.~\eqref{eq:subleadingGreenfun}]. Considering the structure of the subleading braiding factors,
$M_{++}=M_{--}=-M_{+-}=-M_{-+}$, the tunneling rates are proportional to
\begin{subequations} 
    \begin{align} \label{eq:tunnelingratesub_n_1}
W_{\text{m}\rightarrow \text{d}}^{\textrm{n,sub} } \propto & \Big(\frac{\sum_{p_{\textrm{I}}=\pm 1}|\xi_{\textrm{iQPC},(p_{\textrm{M}}=+1,p_{\textrm{I}})}|^{2}}{\sum_{\mathbf{p}} |\xi_{\textrm{iQPC},\mathbf{p}}|^{2} }- \frac{\sum_{p_{\textrm{I}}=\pm}|\xi_{\textrm{iQPC},(p_{\textrm{M}}=-1,p_{\textrm{I}})}|^{2}}{\sum_{\mathbf{p}} |\xi_{\textrm{iQPC},\mathbf{p}}|^{2} } \Big) \nonumber \\
        \times & \Big(\frac{\sum_{p_{\textrm{D}}=\pm} |\xi_{\textrm{dQPC},(p_{\textrm{D}},p_{\textrm{M}}=+1)}|^{2} }{ \sum_{\mathbf{p}} |\xi_{\textrm{dQPC},\mathbf{p}}|^{2} } -\frac{\sum_{p_{\textrm{D}} = \pm}   |\xi_{\textrm{dQPC},(p_{\textrm{D}},p_{\textrm{M}}=-1)}|^{2} }{ \sum_{\mathbf{p}} |\xi_{\textrm{dQPC},\mathbf{p}}|^{2} }\Big),  \quad \textrm{for the A-Pf} \\
W_{\text{d}\rightarrow \text{m}}^{\textrm{n,sub}} \propto & \Big(\frac{\sum_{p_{\textrm{I}}=\pm 1}|\xi_{\textrm{iQPC},(p_{\textrm{M}}=+1,p_{\textrm{I}})}|^{2}}{\sum_{\mathbf{p}} |\xi_{\textrm{iQPC},\mathbf{p}}|^{2} }- \frac{\sum_{p_{\textrm{I}}=\pm}|\xi_{\textrm{iQPC},(p_{\textrm{M}}=-1,p_{\textrm{I}})}|^{2}}{\sum_{\mathbf{p}} |\xi_{\textrm{iQPC},\mathbf{p}}|^{2} } \Big) \nonumber \\
        \times & \Big(\frac{\sum_{p_{\textrm{D}}=\pm} |\xi_{\textrm{dQPC},(p_{\textrm{D}},p_{\textrm{M}}=+1)}|^{2} }{ \sum_{\mathbf{p}} |\xi_{\textrm{dQPC},\mathbf{p}}|^{2} } -\frac{\sum_{p_{\textrm{D}} = \pm}   |\xi_{\textrm{dQPC},(p_{\textrm{D}},p_{\textrm{M}}=-1)}|^{2} }{ \sum_{\mathbf{p}} |\xi_{\textrm{dQPC},\mathbf{p}}|^{2} }\Big), \quad \textrm{for the A-Pf}.
        \label{eq:tunnelingratesub_n_2}
    \end{align}
\end{subequations}
These tunneling rates also vanish when either of the symmetry conditions in Eq.~\eqref{eq:vanishingcurrentcond} is satisfied. Under either condition, the contributions from different neutral-type sectors cancel pairwise within each tunneling rate. This cancellation originates from the structure of the subleading braiding factors,
$M_{++}=M_{--}=-M_{+-}=-M_{-+}$. 
Thus, processes involving the same neutral types and those involving opposite types enter with opposite signs. Under the symmetry conditions, these opposite-sign contributions are weighted equally and cancel in the tunneling rates. Consequently, the subleading braiding process does not contribute to the transport observables, including the neutral Fano factor $\mathcal{F}_{\text{neutral}}$, for the A-Pf edge. 
As discussed earlier, our experimental observation of a vanishing tunneling current at the detection QPC in the neutral-anyon configuration implies that one of these symmetry conditions is satisfied. Accordingly, for the A-Pf edge, the subleading braiding process does not contribute to transport in the neutral-anyon configuration realized in our experiments.

\section{Incoherent heating model}

We examined whether a purely thermal mechanism (which we call an incoherent heating model) could account for the neutral Fano factor observed in our experiments for the neutral-anyon configuration, assuming the possibility that the edges of our devices did not reach the ideal regime described by the fixed-point Hamiltonian [Eqs.~\eqref{PHPfH} and \eqref{APfH} for the PH-Pf and A-Pf states, respectively] at the measurement temperature of 10 mK in the presence of the source voltage $V_\text{S}$. In this heating scenario, QP tunneling at the injection QPC generates a non-equilibrium distribution on the middle edge, which relaxes locally into an elevated effective temperature of the upstream neutral sector. The heated neutral mode then propagates to the detection QPC, and contributes to $S_{\text{D}1}$ at drain D1. For this heating scenario, we developed a model following Refs.~\cite{Park2019, Spanslatt2020}. Using this model, we computed the neutral Fano factor $\mathcal{F}_{\text{neutral}}$, and found that this model cannot explain the experimental data in Fig.~3(b).

This incoherent heating model involves several length scales. The first length scale is the inter-QPC separation $L =1.4 \,\, \mu\text{m}$ in our experiments. The second is the charge-equilibration length, $l_{\textrm{ch-eq}}$. Over this length scale, the inter-mode charge tunneling establishes the same voltage of the modes, which arises when the experimental voltage or temperature cuts off the renormalization-group flow before the edges reach the fixed-point Hamiltonian
[Eqs.~\eqref{PHPfH} and \eqref{APfH} for the PH-Pf and A-Pf states, respectively].
This equilibration process leads to a voltage drop on the upstream side, generating a Joule heating. Such local charge relaxation can also be relevant for the PH-Pf edge, even though its minimal low-energy model contains only a single charge mode, because additional charge degrees of freedom generated by particle-hole conjugation (which may be operative at a finite temperature) participate in short-distance equilibration; see Ref.~\cite{Heiblum2020} for more details. 
The third is the thermal equilibration legnth $l_{\text{th-eq}}$. Over this length scale,
energy exchange among the edge modes leads to thermal equilibration, which also arises when the experimental voltage or temperature cuts off the renormalization-group flow before the edge reaches the fixed-point Hamiltonian
[Eqs.~\eqref{PHPfH} and \eqref{APfH}]. In that case, residual non-fixed-point terms provide a finite energy exchange between edge modes, resulting in a finite thermal-equilibration length~\cite{Ma2020}. 
The last is the energy-loss length $l_{\text{loss}}$, over which energy can be lost from the edge to the FQH bulk or to other environmental degrees of freedom. 

Our heating model assumes the following hierarchy of the length scales:
\begin{align} \label{eq:lengthhierarchy}
    l_{\text{ch-eq}} \ll L \ll l_{\text{loss}} \ll l_{\text{th-eq}}\,.  
\end{align}
The first inequality, $l_{\text{ch-eq}}\ll L$, implies that the charge distribution produced by tunneling at the injection QPC relaxes locally before reaching the detection QPC. Therefore, the neutral excitation arriving at the detection QPC is effectively characterized by an elevated temperature. The second inequality, $L\ll l_{\text{loss}}$, ensures that the excess heat generated near the injection QPC can propagate to the detection QPC without being substantially lost to the bulk or to the environment. Finally, $l_{\text{loss}}\ll l_{\text{th-eq}}$ means that heat loss occurs before the edge modes establish a full thermal equilibration 
with a common temperature among the edge modes. In this regime, the detection QPC probes a locally heated upstream neutral sector, while the incoming charge sector remains cold~\cite{Ma2020}.

The assumption of $L \gg l_{\text{ch-eq}}$ should be contrasted with the opposite coherent limit of $L \ll l_{\text{ch-eq}}$, in which
the time-domain braiding process is operative. 
The remaining hierarchy, $L \ll l_{\text{loss}} \ll l_{\text{th-eq}}$, is motivated by experimental observations. Absent thermal equilibration over relatively very long distances was experimentally observed in both GaAs~\cite{Melcer2022} and graphene~\cite{Srivastav2022, Kumar2022} FQH devices. 
In particular, Ref.~\cite{Melcer2022} measured 
a quantized thermal conductance over distance $200 \, \mu$m, consistent with a picture of negligible thermal equilibration, i.e., $l_{\text{th-eq}} \gg 200 \, \mu$m. However, the upstream-noise measurements in the same device show an exponential decay with a characteristic length of order $l_{\text{loss}}\approx 200\,\mu$m, which was attributed to dissipation of energy to the environment. The separation between charge and thermal equilibration length scales, $\ell_{\rm ch\text{-}eq}\ll \ell_{\rm th\text{-}eq}$, is also supported by the graphene measurements and the accompanying chiral-Luttinger-liquid analysis of Ref.~\cite{Srivastav2022,Kumar2022}.
These observations make the hierarchy $L \ll l_{\text{loss}} \ll l_{\text{th-eq}}$ a plausible phenomenological regime for the present experiment.

The heating model is below described on a more quantitative level for the neutral-anyon configuration. The DC voltage $V_{\textrm{S}}$ is applied to source S1, and the resulting noise is measured in drain D1. The charge current injected into the middle edge through QPC$1$ is written as
\begin{align}
I_{{\rm iQPC}}(V_{\textrm{S}})
=
G_0\int_0^{V_{\textrm{S}}} dV'
R_{{\rm iQPC}}^{\rm diff}(V') ,
\end{align}
where $R_{{\rm iQPC}}^{\rm diff}(V')$ is the differential reflection probability, and $G_0 = e^2 /(2h)$ is the conductance of the active fractional edge modes. Charge conservation then fixes the effective dc voltage of the middle edge,
\begin{align}
V_{\rm eff}
=
\frac{I_{{\rm iQPC}}(V_{\textrm{S}})}{G_0}
=
\int_0^{V_{\textrm{S}}} dV'
R_{{\rm iQPC}}^{\rm diff}(V')
\equiv
V_{\textrm{S}} R_{{\rm iQPC}} .
\end{align}
The total energy current injected into the middle edge M is
\begin{align}
J^E_{\text{M}}
=
G_0\int_0^{V_{\textrm{S}}} dV'
V' R_{{\rm iQPC}}^{\rm diff}(V') .
\end{align}
Only the part of this energy current in excess of the DC contribution is available for local heating and the generated power reads 
\begin{align} \label{eq:jouleheatingpower}
P_{\rm heat}
=
G_0\int_0^{V_{\textrm{S}}} dV'
V' R_{{\rm iQPC}}^{\rm diff}(V')
-
\frac{1}{2}G_0V_{\rm eff}^2 .
\end{align}
Assuming that this generated heat is fully used for heating the edge modes and 
the heat is redistributed among the edge modes by the ratio of the respective central charge, we found that the elevated temperature $T_{\text{ex}}$ of the modes satisfies
\begin{align} \label{eq:heatpower}
\frac{\pi^2 k_\text{B}^2 }{6h} (|c_{n}|+|c_{c}|)(T_{\text{ex}}^2-T^2)
=
P_{\rm heat}\,,
\end{align}
where $T$ is the ambient temperature of the system. $c_{c}$ and $c_{n}$ the central charges of the charged and neutral modes, respectively.

The neutral mode leaving the hot spot then carries this effective temperature $T_{\text{ex}}$ toward the detection QPC (QPC$2$). At the same time, the charge mode impinges on the QPC with the ambient temperature $T$. The elevated temperature of the neutral mode modifies the tunneling rates at the detection QPC and therefore affects both the tunneling noise $S_{\rm dQPC}$ and the differential response $\partial I_{\rm dQPC}/\partial V_{\rm vir}$ as 
\begin{align}
    S_{\text{D}1}^{\text{exc}} = S_{\text{dQPC}}(V_{\textrm{S}}) - 4 k_\text{B} T \left. \frac{\partial I_{\text{dQPC}}(V_{\textrm{S}})}{\partial V_{\text{vir}}} \right|_{V_{\text{vir}} = 0} \,. 
\end{align}
To quantify how much excess noise is generated, we adopt the Luttinger liquid theory, in which the effect of elevated temperature for the neutral mode is taken into account through the equilibrium Green function for the edge $M$: 
\begin{align}
    G^{\lessgtr}_{\text{M}} (t) = \frac{\pm i}{2\pi a} \Big(\frac{\pi k_\text{B} T a v_c^{-1}}{\sin[\pi k_\text{B} T (a v_c^{-1} \mp i t)]} \Big)^{\delta_c }  \Big(\frac{\pi k_\text{B} T_{\text{ex}} a v_n^{-1}}{\sin[\pi k_\text{B} T_{\text{ex}} (a v_n^{-1} \mp i t)]} \Big)^{\delta_n } 
\end{align}
In contrast, the Green function of the detection edge is in equilibrium at temperature $T$. Plugging these Green functions into the expressions for the tunneling rates in Eq.~\eqref{eq:tunnelingratesQPC}, we computed the
noise $S_{\text{D}_1}^{\text{exc}}$, the differential reflection probability $R_{\text{dQPC}}^{\text{diff}}$, and hence the neutral Fano factor $\mathcal{F}_{\text{neutral}}$. By expanding the Green functions in powers of $\Delta T / T$ with $\Delta T \equiv T_{\text{ex}} - T$ and integrating the resulting linear-order contribution over time, we found the linear relation  
\begin{align}
    \label{eq:heatingnoiseRefrelation}
S_{\text{D}1}^{\text{exc}} \simeq 2 G_0 R_{\text{dQPC}}^{\text{diff}} \beta  k_\text{B} \Delta T\,, \qquad \text{with } \beta = \frac{\delta_n}{\delta_c + \delta_n}\,.
\end{align}
The factor $\beta = \delta_n/(\delta_c+\delta_n)$ reflects that only the neutral sector of the tunneling operator is heated, while the charge sector remains at the ambient temperature.
Using $I_{{i\rm QPC}}=G_0 V_\text{S} R_{{\rm iQPC}}$ together with Eqs.~\eqref{eq:heatingnoiseRefrelation} and \eqref{eq:heatpower}, we found the expression for the neutral Fano factor 
\begin{align} \label{eq:Fanoneutralheatingformula}
    \mathcal{F}_{\rm neutral}
    =
    \frac{S_{\rm D1}^{\rm exc}}
    {2e^* I_{{\rm iQPC}} R_{{\rm dQPC}}^{\rm diff}}
    =
    \frac{1}{e^* V_\text{S} R_{i{\rm QPC}}}
    \frac{\delta_n}{\delta_c+\delta_n}
    \left[
    \sqrt{
    (k_\text{B} T)^2+
    \frac{6hP_{\rm heat}}
    {\pi^2 (|c_c|+|c_n|)}
    }
    - k_\text{B} T
    \right]\,. 
\end{align}
This expression $\mathcal{F}_{\text{neutral}}$ holds for both PH-Pf and A-Pf edges, provided that $\delta_c$, $\delta_n$, $c_c$, and $c_n$ are plugged in accordingly. For the PH-Pf edge, $\delta_c = 1/8$, $\delta_n = 1/8$, $|c_c| = 1$, and $|c_n|= 1/2$, and for the A-Pf edge, $\delta_c = 1/8$, $\delta_n =3/8$, $|c_c| = 1$, and $|c_n| = 3/2$. 

\begin{figure}[t]
	\centering
	\includegraphics[width = .7\textwidth]{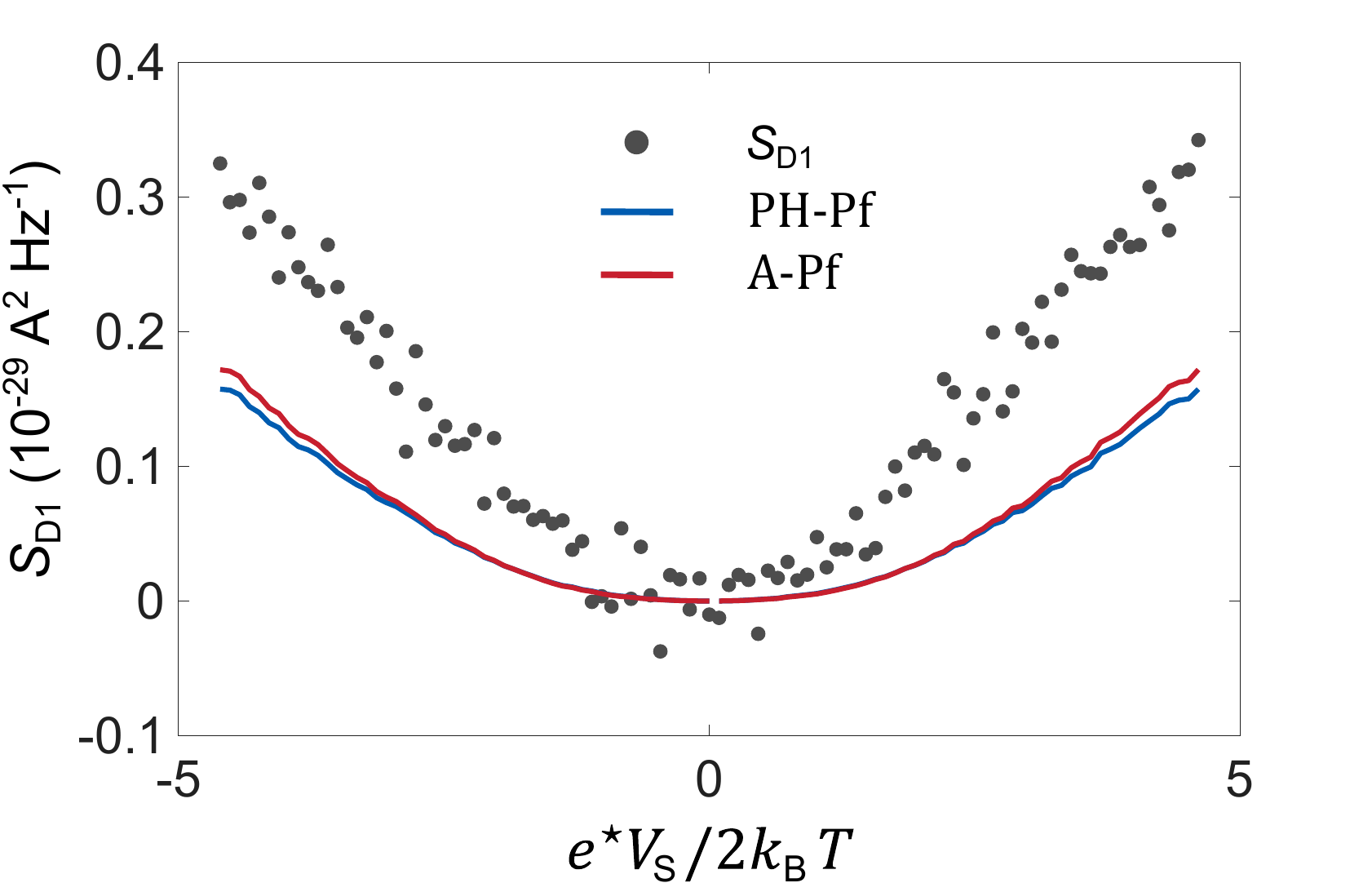}
	\caption{Excess noise $S_{\text{D}1}$ as a function of $e^* V_{\textrm{S}}/(2 k_\text{B} T)$ for the neutral-anyon configuration. Gray dots denote the experimental data from data set 4, identical to those shown in panel (g) of Fig.~\ref{fig:S4}. The blue and red curves show the theoretical predictions obtained from the incoherent heating model for the PH-Pf and A-Pf edges, respectively. For both edges, the heating model underestimates the measured excess noise by roughly a factor of two over the relevant voltage range. This discrepancy indicates that the incoherent heating of the neutral sector alone cannot account for the observed excess noise $S_{\text{D}1}$.} 
	\label{fig:heating}
\end{figure}

Employing Eq.~\eqref{eq:Fanoneutralheatingformula} with the heat power $P_{\text{heat}}$ [Eq.~\eqref{eq:jouleheatingpower}] evaluated using the experimentally measured differential reflection probabilities at the injection QPC, we computed the excess noise $S_{\text{D1}}^{\text{exc}}$
and plotted it 
as a function of $e^*V_{\text{S}}/(2 k_\text{B} T)$ in Fig.~\ref{fig:heating}. The heating-model predictions for both PH-Pf and A-Pf edges are substantially smaller than the experimental data presented in Fig.~3 of the main text and the Supplementary Figs.~\ref{fig:S1}-\ref{fig:S4}. This discrepancy indicates that the observed excess noise cannot be explained by the incoherent heating scenario.

\end{document}